\documentclass[preprint,showpacs,preprintnumbers,amsmath,amssymb,
nofootinbib,eqsecnum,12pt]{revtex4}

\usepackage{graphicx}

\newlength{\dummysp}
\newcommand{\tr}{\mathop{{\hbox{Tr} \, }}\nolimits}

\newcommand{\half}{\frac{1}{2}}

\newcommand{\beq}{\begin{eqnarray}}
\newcommand{\eeq}{\end{eqnarray}}
\newcommand{\nnn}{ \nonumber \\ }
\newcommand{\p}{{\partial}}
\newcommand{\e}{{\epsilon}}
\newcommand{\s}{{\sigma}}

\newcommand{\ord}[1]{{{\cal O}(#1)}}
\newcommand{\gappeq}{\mathrel{\rlap {\raise.5ex\hbox{$>$}}
{\lower.5ex\hbox{$\sim$}}}}
\newcommand{\lappeq}{\mathrel{\rlap{\raise.5ex\hbox{$<$}}
{\lower.5ex\hbox{$\sim$}}}}
\newcommand{\myref}[1]{(\ref{#1})}

\newcommand{\ben}{\begin{enumerate}}
\newcommand{\een}{\end{enumerate}}
\newcommand{\bit}{\begin{itemize}}
\newcommand{\eit}{\end{itemize}}

\newcommand{\Ncal}{{\cal N}}

\newcommand{\Ub}{{\bar U}}
\newcommand{\FUb}{F_\Ub}

\newcommand{\susy}{SUSY}
\newcommand{\Susy}{SUSY}

\newcommand{\csy}{SUSY}
\newcommand{\zIR}{z_1}
\newcommand{\zir}{z_1}
\newcommand{\adsv}{\text{AdS}_5}
\newcommand{\adsvt}{$\adsv$}
\newcommand{\hc}{\text{h.c.}}
\newcommand{\MET}{\not \hspace{-3pt} E_T}
\newcommand{\METa}{\not \hspace{-4pt} E_T}
\newcommand{\METaa}{\hbox{$\not \hspace{-3pt} E_T$}}

\newcommand{\qbar}{{\bar q}}
\newcommand{\enot}[2]{$#1 \times 10^{#2}$}

\newcommand{\psib}{{\bar \psi}}
\newcommand{\dpmet}{$2\gamma+\METaa$}
\newcommand{\Lamlt}{{\tilde m}_{1,2}}

\def\[{\left [}
\def\]{\right ]}
\def\({\left (}
\def\){\right )}

\begin{document}

\noindent

\hfill UMN-TH-2542/07

\hfill FTPI-MINN-07/09

\vskip 1cm

\begin{center}
{\bf \Large A Gravity Dual and LHC Study of \\ 
\vskip 0.2cm Single-Sector Supersymmetry Breaking}
\end{center}

\vspace{0.02cm}
\begin{center}
{\sc Maxime Gabella,\footnote{{\tt gabella@cern.ch}}$^{a,b}$} 
{\sc Tony Gherghetta\footnote{{\tt tgher@physics.umn.edu}}$^{b}$} 
{\small and}
{\sc Joel Giedt\footnote{{\tt giedt@physics.umn.edu}}$^{c}$}
\end{center}

\begin{center}

$^a${\it\small ITP, \'Ecole Polytechnique F\'ed\'erale de Lausanne, CH-1015 Lausanne,
Switzerland}

$^b${\it\small School of Physics and Astronomy, 
University of Minnesota, Minneapolis, MN 55455, USA}

$^c${\it\small  William I. Fine Theoretical Physics Institute,
University of Minnesota, \\ Minneapolis, MN 55455, USA}

\end{center}

\vspace{0.2cm}

\begin{abstract}
We propose a gravitational dual of ``single-sector'' models of supersymmetry breaking 
which contain no messenger sector and naturally explain the scale of supersymmetry breaking 
and the fermion mass hierarchy. In five dimensions these models 
can be given a simple interpretation.
Inspired by flux-background solutions of type IIB supergravity, 
a metric background that deviates from AdS$_5$ in the IR breaks 
supersymmetry, while the fermion mass hierarchy results 
from the wavefunction overlap of bulk fermions with a UV-confined 
Higgs field. The first and second generation squarks 
and sleptons, which are localized near the IR brane, 
directly feel the supersymmetry breaking and obtain 
masses of order 10 TeV. These are interpreted as 
composite states of the dual 4D theory. 
The gauginos and third generation squarks and sleptons are
elementary states that obtain soft masses of order 1 TeV at the 
loop level via direct gauge mediation.
This particle spectrum leads to distinctive signatures at 
the LHC, similar to the usual gauge mediation 
with a neutralino NLSP that decays promptly to a 
gravitino LSP, but with lower event rates. 
Nevertheless we show that with 1-10 fb${}^{-1}$ of LHC data
``single-sector" models can easily
be detected above background
and distinguished from conventional gravity and gauge
mediation.

\end{abstract}



\maketitle

\section{Introduction}
With the advent of reliable techniques to
study dynamical supersymmetry breaking~(DSB)~\cite{DSBrvw}
in the mid-1990's, gauge mediation\footnote{
For a review of gauge mediation, and references to the original articles,
see \cite{Giudice:1998bp}.}
flourished as an alternative
to its elder sibling, supergravity mediation.
Gauge mediation solves the flavor
problems of supergravity mediation \cite{sugfp}, due to
universality at a relatively low scale.
At the same time, several proposals \cite{Arkani-Hamed:1997fq,
Luty:1998vr,mmssm} were made to combine the nonperturbative dynamics
that broke supersymmetry (\csy) with a modicum of
compositeness in the Standard Model sector
of the theory.  These models were closely
related to ideas of MSSM-compositeness \cite{scomp}.
As in older non-\csy\ models \cite{cfl},
light fermion masses could be explained on
dimensional grounds.  On the other hand, the scalar partners of these
light fermions obtained masses much larger than the
electroweak scale, because they coupled strongly to the DSB sector and
were not protected by chiral symmetries.
In the examples that were constructed during
this period, the quantitative details of
the multi-TeV-scale nonperturbative dynamics
was ``incalculable'',
signifying qualitative but not especially quantitative predictions.

Recently, gauge/gravity duality ideas based on
the AdS/CFT correspondence  in type IIB string theory~\cite{ador}
have been used to give a four-dimensional (4D) holographic description for
models in a warped extra dimension~\cite{wpac}.
This has led to the remarkable result 
that strongly coupled 4D gauge
dynamics can be modeled with a five-dimensional (5D),
weakly coupled gravitational theory.  
In this approach, classical field theory
computations are able to capture the dominant
effects of the strongly coupled 4D theory.  
For example, the Randall-Sundrum
model~\cite{Randall:1999ee} can be given a purely 
holographic interpretation as a 4D
composite Higgs model. The warp factor is used to obtain 
a low symmetry breaking scale which is then identified as the dynamical 
electroweak symmetry 
breaking scale. In fact more recent composite Higgs models consistent with 
all electroweak precision tests have been constructed primarily
motivated from the gravity side~\cite{chm}.
Ideas from AdS/CFT have even been applied
to QCD, where the chiral symmetry breaking scale
is related to the warp factor~\cite{ADSQCD}.
Similarly warped extra dimension models 
have been used to break \susy~\cite{wmssm}, where the
warp factor is now used to generate a low SUSY-breaking
scale which is then identified as a DSB scale. In these models 
boundary conditions were used to break 
\susy. In the present work, we pursue this idea of 
relating the warp factor with a dynamically generated scale
in the context of realistic, strongly coupled
4D \Susy\ gauge theories, softly broken by the
effects of DSB.  A simple 5D gravitational
dual will be described that will allow previously 
``incalculable" particle mass spectra
to be calculated.

A key insight is gained from recent
developments in string/M-theory
and its effective supergravity descriptions.
We note that the realizations of the AdS/CFT correspondence 
that are on a firm theoretical footing arise in just this context.  
Whereas the original correspondence was formulated
in models with maximal
\susy\ \cite{ador}, it has been known for some time
now how to construct rigorous gauge/gravity dual systems
with far less \susy\ \cite{Klebanov:1998hh,
Klebanov:2000hb}.
In particular, the Klebanov-Strassler
solution \cite{Klebanov:2000hb} preserves
$\Ncal=1$ \susy\ in the 4D gauge theory.
In recent work \cite{Gherghetta:2006yq},
the realization of bulk scalar modes 
was studied in this type IIB supergravity context,
and related to scalars that would appear
in \adsvt\ phenomenological models. However,
for realistic models, \susy\ must be spontaneously
broken and it is natural to ask whether there are
non-\susy\ supergravity backgrounds that might
be useful for this purpose. 

In fact there exist deformations of the Klebanov-Strassler
background that softly break
all of the supersymmetry in the infrared (IR)
of the 4D theory \cite{Borokhov:2002fm,Kuperstein:2003yt}.  It will be shown that
when the ten-dimensional (10D)
supergravity theory is reduced to five dimensions, the
background geometry is a deformation of \adsvt,
with the modification growing stronger as
one moves further into the AdS throat---corresponding
to \susy-breaking near the ``IR brane'' 
of phenomenological models.  Consequently we will use
the non-\Susy\ background of \cite{Kuperstein:2003yt} as a starting point
to construct a string-inspired model with
realistic phenomenology.  As in the Randall-Sundrum 
model~\cite{Randall:1999ee} we will introduce both a  ``UV brane'' 
and an ``IR brane", and then consider bulk fields~\cite{gp} by 
embedding the MSSM into 
a slice of the deformed \adsvt, with all but
the Higgs fields propagating in the bulk.
At low energies, the theory is described by the
MSSM with \susy-breaking soft terms that are
determined by the deformed \adsvt\ geometry; since the deformation 
is determined by a single parameter, the model is quite economical.
However the deformed geometry only gives sizable soft masses to
scalar fields localized near the IR brane.  
For the remaining sparticles we show that soft masses of the
requisite scale are generated radiatively. 

In particular since the gaugino masses arise at one loop in the 4D description, they
must have a tree level interpretation in the 5D theory.
However, the naive action in the deformed background
has a $U(1)_R$ symmetry that protects the
gauginos from acquiring a mass. 
As we show, the tree-level mass arises from an additional contribution
to the gaugino action due to the nontrivial, $U(1)_R$-violating
flux background in the underlying type IIB 
supergravity.  

The dual 4D theory that we obtain is remarkably similar 
to a purely 4D ``single-sector" model
constructed in Ref.~\cite{Arkani-Hamed:1997fq}.
We determine the crucial ratio $F/M$ that encodes the messenger
dynamics by comparing scalar masses of
composite states to the results in Ref.~\cite{Arkani-Hamed:1997fq}.
The ratio $F/M$ then determines the perturbative
corrections that provide soft terms
for all fields localized near or at the
UV brane.  Exploiting well-known
results in gauge mediation, we find that
the Higgs fields obtain the necessary soft masses 
from this effect, rendering electroweak symmetry-breaking
viable.

\begin{figure}
\begin{center}
\includegraphics[width=3in,height=3in]{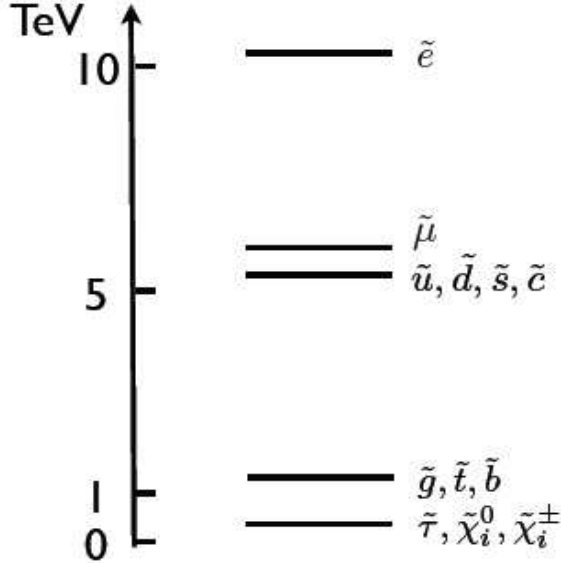}
\caption{The generic mass spectrum of the 5D gravity model showing
the heavy first and second generation scalars and lighter third
generation scalars, gluinos, neutralinos and charginos. The LSP is the gravitino 
(not shown).
\label{genmass}}
\end{center}
\end{figure}

Consequently the particle spectrum in our model has very distinctive 
features and a generic spectrum is shown in Fig.~\ref{genmass}.
The first and second generation of scalar partners are
very heavy. These large masses do not destabilize the 
Higgs mass via radiative corrections
because of (1) small Yukawa couplings and (2)
degeneracies at the messenger scale that prevent
large one-loop hypercharge Fayet-Iliopoulos (FI) terms.\footnote{Here
it is important that the messenger scale $M$ in our model is $\ord{100}$ TeV,
so that splittings that would disturb the degeneracies
are not introduced under renormalization group evolution
to the electroweak scale.}
As will be discussed, these degeneracies are
also necessary in order to satisfy flavor changing
neutral current (FCNC) constraints.
This spectrum is similar to that considered in 
Refs.~\cite{Dimopoulos:1995mi, Pomarol:1995xc} and is
also reminiscent of the ``more minimal" 
supersymmetric standard model~\cite{mmssm}, for which heavy 
first two generation scalar fields were considered to ameliorate flavor problems.
The LSP is the gravitino, which means that in our model the lightest
neutralino, $\tilde \chi_1^0$, is the NLSP.  Because
the messenger scale is relatively low, the decay length
of $\tilde \chi_1^0$ is less than 1 mm.
This leads to a \dpmet\ (two hard photons
and missing transverse energy)
signal at the LHC.  Although the event rate 
is reduced ($\sim$ 50\%) compared to conventional
gauge-mediated SUSY breaking (GMSB) due to the heavy first and
second generation squarks and sleptons being inaccessible at the LHC,
we show that this signal can easily be seen
with 1-10 fb$^{-1}$ integrated luminosity.

The plan for the rest of the paper is as follows:
In Section \ref{msu} we present the phenomenological
model from the 5D point of view and show how a natural 
supersymmetry breaking scale  and a fermion mass hierarchy
can be simply explained by a warped extra dimension. 
In Section \ref{4dd} we describe the dual 4D holographic
interpretation of our model and its relation to 4D single-sector models.
We also discuss gauge-mediated contributions to soft masses 
and address important phenomenological constraints from FCNC's and
naturalness.  In Section \ref{lhc} we discuss in detail the LHC \dpmet\
signal of our model.  We compare to rates in 
conventional GMSB, and illustrate how simple cuts allow for the
removal of virtually all Standard Model (SM) 
background. Finally, in the Appendices we present details of 
bulk scalar fields, fermions and Yukawa interactions in the deformed 
warped geometry as well as show how our 5D model is inspired from
the 10D type IIB supergravity solution of \cite{Kuperstein:2003yt}.

\section{The 5D gravity model}
\label{msu}
We begin by defining our model using a geometrical 5D framework. In this way 
a naturally small scale of supersymmetry breaking with fermion mass hierarchies
will be manifest. This leads to a characteristic superparticle mass spectrum with features
that depend only on the broad properties of the underlying geometry. 
However the advantage of the 5D model is that the particle mass spectrum can be
reliably calculated, and  we will also explicitly present a particular mass spectrum
of our 5D model to illustrate this calculational ability.

\subsection{Deformed \adsvt}
In order to break supersymmetry we will consider an
effective 5D model that is motivated from a 10D type IIB supergravity 
solution \cite{Kuperstein:2003yt}, obtained by perturbing
the well-known Klebanov-Strassler supersymmetric background \cite{Klebanov:2000hb},
using techniques developed in \cite{Borokhov:2002fm}. 
In Appendix~\ref{sugra}, we describe in detail how
the effective deformed nonsupersymmetric 5D
background metric is obtained from a dimensional
reduction of the 10D metric.\footnote{The flux
background is also deformed, and is relevant
to the gaugino masses, as will be discussed.}
The resulting 5D geometry will be parametrized as:
\beq
&& ds^2 = A^2(z) \( -dt^2 + d {\vec x}^2 + dz^2 \)~, \nnn
&& A^2(z) = \frac{1}{(k z)^2} \[1 - \e \(\frac{z}{z_1} \)^4 \],
\label{defmetric}
\eeq
where $k$ is the AdS curvature scale, and $z_0 \leq z \leq z_1$ with 
$z_0$, $z_1$ the positions of the UV and IR branes respectively.
The parameter $\e$ is related to variables in the original 10D solution 
(see Appendix~\ref{sugra}), although for our phenomenological 
purposes we only need assume 
it to be an arbitrary but small, positive parameter.
The $\e \to 0$ limit is just a slice of AdS$_5$, which is the
5D background setup used in the Randall-Sundrum 
model~\cite{Randall:1999ee}.

It is clear that the deformation of  AdS$_5$  dominates
in the IR, where $z$ is the largest.  Physical arguments and the
fact that this is meant to be a small perturbation with
\susy\ breaking far below the AdS curvature scale $k \sim m_P$, where $m_P$ is the
(reduced) Planck scale, requires that
\beq
\e < 1,  \quad k \zIR \lappeq \frac{m_P}{\text{TeV}}\simeq 10^{15}.
\eeq
The actual value of the IR scale $z_1^{-1}$ will be determined later by satisfying
constraints from FCNC's and naturalness.

\subsection{MSSM in the bulk}
Next we introduce the MSSM field content into the bulk with metric \myref{defmetric}.  
In the supersymmetric limit ($\e \to 0$) these 5D fields
propagate in a slice of AdS and satisfy nontrivial boundary
conditions~\cite{gp}. Upon compactification to four
dimensions, the massless zero modes of the Kaluza-Klein 
towers are identified with the 4D MSSM fields. 
Unlike in the Randall-Sundrum model,
the warp factor is used to set the scale of supersymmetry breaking and 
is parametrized by the deformation of the AdS metric (\ref{defmetric}).
Consequently the Higgs fields need not be localized on the IR brane and in fact we
assume them to be confined on the UV brane where their masses are protected by supersymmetry.  
Instead the IR brane is the source of supersymmetry breaking in our model.

Furthermore, the extra dimension is used to naturally generate small Yukawa couplings
for the massless fermions by wavefunction overlap with the UV-confined Higgs field. In particular
this means that the first two fermion generations are localized predominantly near the IR brane,
while the third generation fermions are nearer to the UV brane. In this way the warped 
extra dimension not only helps to explain the scale of supersymmetry breaking but also the 
fermion mass hierarchies.\footnote{A Higgs
localized on the IR brane can also be considered.  
However, the UV-localized scalar superpartners
of the first two generations only feel gauge mediation.
For them to obtain sufficiently large soft masses, the scale of the
IR brane must be $\ord{100}$ TeV.  This will also be the scale of
the stop mass, since it is IR localized in this scenario.
Such a large stop mass would clearly destabilize the Higgs mass
through its $\ord{1}$ Yukawa coupling.
Nevertheless, these problems are not insurmountable, but would 
require additional assumptions on the model.}

\subsection{Fermion masses}
Let us first consider the SM fermions.  As shown in
Appendix \ref{fms}, each SM fermion is embedded 
into its own 5D field.  The zero mode profile for each fermion $i$ 
is given by
\beq
f_i(z) \propto z^{\frac{1}{2}-c_i}~,
\eeq 
where the exponent depends on a bulk mass parameter 
$c_i$. For $c_i> 1/2~(c_i<1/2)$ the zero mode is localized 
near the UV (IR) brane.  The wavefunction
overlap of the fermion zero modes with the UV-confined Higgs fields  $z_*=z_0=1/k$, 
using the expression (\ref{eq:Y(c)}) in Appendix~\ref{ycsect}, leads to the 4D Yukawa couplings
\beq
Y_{\psi} = Y^{5D}_{\psi} k \sqrt{\frac{1/2-c_L}{(kz_1)^{1-2c_L}-1}}\sqrt{\frac{1/2+c_R}
{(kz_1)^{1+2c_R}-1}}~.
\eeq
This expression is used to solve for the $c$ parameters using the values of the 4D Yukawa couplings
and assuming $10^{-3} \lesssim Y^{5D}_{\psi} k \lesssim 1$. 
(It will be seen below that it is necessary to allow a small
hierarchy here, in order to avoid FCNC's from the squarks.
Essentially, the $c$'s must be degenerate among first and
second generation quarks in order for the corresponding
squarks to be degenerate.)
The results are listed in Table~\ref{SMfields}.
\begin{table}
\begin{center}
\begin{tabular}{|c||c|c|c|c||c||c|c|c|c||c||c|c|c|c|}
\hline
& $\bar m(m_Z)$  & $c_L$ & $-c_R$ & $Y^{5D}$ & 
& $\bar m(m_Z)$  & $c_L$ &  $-c_R$ & $Y^{5D}$ &  
& $\bar m(m_Z)$  & $c_L$ & $-c_R$ & $Y^{5D}$ \\
\hline\hline
$e$ & 0.503 MeV & 0.350 & 0.350 & 1 
& $d$ & 3.9 MeV & 0.456 & 0.456 & 0.059 
& $u$ & 1.7 MeV & 0.456 & 0.456 & 0.0025 \\
\hline
$\mu$ & 103.9 MeV & 0.467 & 0.467 & 1
& $s$ & 67.6 MeV & 0.456 & 0.456 & 1
& $c$ & 0.58 GeV &0.456 & 0.456 & 0.849 \\
\hline
$\tau$ & 1.75 GeV & 0.601 & 0.601 & 1
& $b$ & 2.9 GeV & 0.69 & 0.648 & 1
& $t$ & 166 GeV & 0.69 & 5.341 & 1 \\
\hline
\end{tabular}\caption{\small Standard Model $\overline{\text{MS}}$
running fermion masses 
at the scale $m_Z$, as determined by Softsusy~\cite{Allanach:2001kg}.
Also shown are the corresponding $c$ values and
5D Yukawa couplings (in units of $k$) for the case of
UV Higgses and $\tan\beta = 10$.}
\label{SMfields}
\end{center}
\end{table}

Indeed it is seen from these values
that the lighter generations are closer to the IR brane while the third generation is UV-localized.
Since each SM fermion is contained in a chiral supermultiplet, the corresponding scalar 
superpartner will be localized at the same place in the supersymmetric limit.
In the deformed case, the scalar localization is qualitatively
unchanged.  This is because the profile is only modified in
the IR, where the deformation is noticeable.

In the supersymmetry breaking background (\ref{defmetric}) it is shown
in Appendix \ref{fms} that the zero modes 
of bulk fermions are not lifted.  The SM fermions are
protected by a chiral symmetry and the gauginos by a U(1)$_R$ symmetry. These symmetries
are both respected by the metric (\ref{defmetric}).  
Massless gauginos are, of course, phenomenologically unacceptable. 
In Appendix \ref{fluxr} we take into account R-symmetry breaking
that arises from the nontrivial flux background that is associated
with the metric \myref{defmetric}, in the type IIB solution
\cite{Kuperstein:2003yt}.  The results that we obtain
accord well with the radiatively generated gaugino mass
that is evident in the 4D dual gauge theory.

\subsection{Scalar masses}
Supersymmetry is broken in the bulk by the IR deformation (\ref{defmetric}) of the AdS metric.
The squarks and sleptons will obtain masses that depend on their localization
in the bulk. However, as mentioned above,
the requirement of obtaining hierarchical Yukawa couplings fixes
the localization of the corresponding scalar superpartners. In this way the scalar
superpartner masses are in fact related to the fermion mass spectrum.

As reviewed in Appendix \ref{sec:ScalarField},
the zero mode profile of a bulk scalar field is given 
at leading order (small corrections are described
in the Appendix) by
\beq
f_i(z) \propto z^{b_i-1}~,
\eeq 
where the exponent depends on a mass parameter $b_i$ of the 5D model.
By supersymmetry~\cite{gp}
\beq
b_i=\frac{3}{2} -c_i,
\label{bcr}
\eeq
which explicitly shows that once the SM fermion localization is set by $c_i$, the 
localization of the scalar zero mode is then fixed.
The values $b_i < 1~ (c_i>1/2)$ correspond to a UV-localized
mode, whereas $b_i > 1~(c_i<1/2)$ is IR-localized.  Clearly
it is the IR-localized scalar modes that are sensitive to the
\susy-breaking background because the deformation is only 
appreciable near the IR brane. 

\begin{figure}
\begin{center}
\includegraphics[width=2.7in,height=3.5in,angle=90]{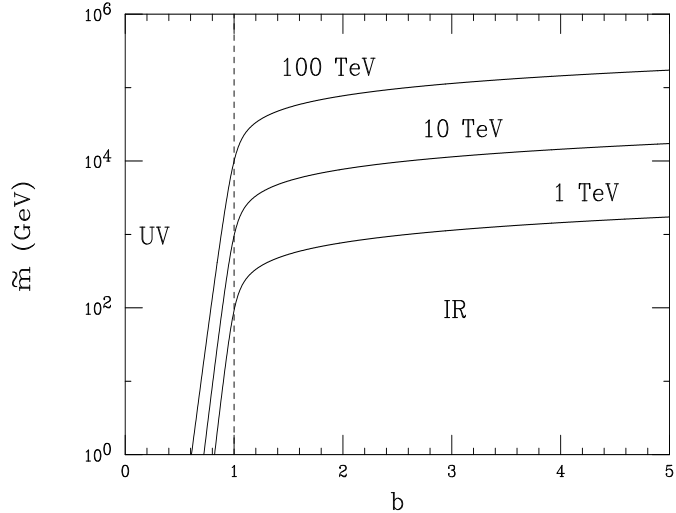}
\caption{The scalar mass-squared  \myref{musf} as a function of $b$ 
for three values of $\zir^{-1} =$ 1, 10 and 100 TeV, with
$\e=0.05$.  Not shown are the exponentially small values at $b \lappeq 1/2$.
The values $b < 1$ correspond to a UV-localized mode, whereas $b > 1$ is IR-localized.
\label{m2f}}
\end{center}
\end{figure}

In the supersymmetry-breaking background (\ref{defmetric}) the scalar zero modes will obtain 
a mass. It is straightforward to analytically solve the equation of motion
for the scalar zero modes by using a linearized (in $\e$) approximation (see Appendix~\ref{sec:ScalarField}).  The scalar mass squared as a function of the localization parameter $b$ 
is given by
\beq
{\widetilde m}^2 = \e \frac{(1-b)(b+10)}{(k\zir)^4} \frac{(k\zir)^{1+b} - (k\zir)^{1-b}}
{ (k\zir)^{1-b} - (k\zir)^{b-1}} k^2 +{\cal O}(\e^2)~.
\label{musf}
\eeq
This expression simplifies in the limit $k\zir \gg 1$.  For $b> 1$ the scalar 
mass simply becomes
\beq
{\widetilde m}\approx \sqrt{\e (b-1)(b+10)} \zir^{-1}~,
\eeq
while for $0<b<1$ we have the approximation:
\beq
{\widetilde m} \approx \sqrt{\e (1-b)(b+10)} (k\zir)^{b-1} \zir^{-1}~.
\eeq
Thus we see that for an IR-localized field ($b>1$) the scalar mass becomes of order the
IR scale $ \zir^{-1}$, while for $b\ll 1$ and $k\zir\sim 10^{13}$
the scalar mass is much less than a GeV. The exact expression (\ref{musf}) is plotted in 
Fig.~\ref{m2f} for three values of $\zir^{-1}=$ 1, 10 and 100 TeV, and for  $\e=0.05$, which
exhibits the above behavior for $b>0$.  
Note that as $b\rightarrow 0$ the coefficient of the
$\e$ term in (\ref{musf}) vanishes and the corresponding mass 
will be given by higher order terms. We have numerically checked that for 
$b<0$ the masses are vanishingly small.

From Eq.~\myref{bcr}, the values of $b_i$ are
determined by the fermion spectrum
of Table~\ref{SMfields}.  We then apply \myref{musf}
to obtain the squark and slepton mass spectrum of Table~\ref{tab:softmUVH}. The AdS curvature 
scale is set by requiring  $m_P^2\simeq M_5^3/k$ where $M_5$ is the 5D Planck scale. Choosing $k\sim 0.1 M_5$ requires $k\simeq 10^{-3/2} m_P=  7.7 \times 10^{16}~ \text{GeV}$. 
Consequently the model parameters are set to
\beq
&& \pi kR = 28.42 , \quad \e = 0.05, \quad \tan\beta=10,\nnn
&& z_0 = k^{-1}, \quad z_1 = (k e^{-\pi k R})^{-1} = (35 ~ \text{TeV})^{-1}.
\label{oset}
\eeq
We see that the first two generations of squarks and sleptons obtain masses of order 
1/10 to 1/20 the Kaluza-Klein mass scale,
\beq
m_{\text{KK}} = \pi z_1^{-1} = 110 ~ \text{TeV},
\label{mkk}
\eeq
but the third generation masses are much smaller. 
As expected since the third generation fermions are near the 
UV brane in order to have a large overlap with the Higgs, the corresponding 
supersymmetry-breaking masses are phenomenologically unacceptable. However 
by considering the dual 4D theory we will show that there is a gauge-mediated contribution 
that gives rise to acceptable third generation squark and slepton masses.

\begin{table}
\begin{center}
\begin{tabular}{||c|c||c|c||}
\hline
Sparticles & $\tilde m$ [TeV] & Sparticles & $\tilde m$ [TeV] \\
\hline
$\tilde e_{L,R}$, $\tilde \nu_{e L}$ & 10.14 &
$\tilde u_{L,R}$, $\tilde d_{L,R}$ & 5.69 \\ \hline
$\tilde \mu_{L,R}$, $\tilde \nu_{\mu L}$ & 5.12 &
$\tilde c_{L,R}$, $\tilde s_{L,R}$ & 5.69 \\ \hline
$\tilde \tau_{L,R}$, $\tilde \nu_{\tau L}$ & 0.468 &
$\tilde b_R$ & 0.149 \\
$\tilde t_L$, $\tilde b_L$ & 0.051 &
$\tilde t_R$ & 0 \\
\hline
\end{tabular}
\caption{\small Soft masses, as determined by \myref{musf}.
The boundary mass parameters $b$ are determined from \myref{bcr},
using the $c$-parameters given in Table \ref{SMfields}.
} 
\label{tab:softmUVH}
\end{center}
\end{table}

\section{The 4D dual model}
\label{4dd}

According to the AdS/CFT correspondence, our 5D phenomenological model in a slice
of AdS admits a 4D dual description in terms of a strongly coupled gauge theory that mixes
with a fundamental sector.  The supersymmetry breaking background (\ref{defmetric}) is 
dual to DSB caused by the strongly coupled gauge theory 
in the IR, with a scale of order 100 TeV. 
The AdS/CFT dictionary identifies UV-localized fields as elementary states
in the fundamental sector and IR-localized fields as bound states of the CFT. Thus our
5D phenomenological model is dual to a supersymmetric fundamental sector
containing the Higgs, third generation fermions and 
gauge fields,  while the first two generation
fermions and sfermions are composite states of the dual gauge theory.

\subsection{Relation to 4D single-sector models}

Interestingly, this dual model is remarkably similar to models constructed purely in
four dimensions.  In particular, the authors of the
``single-sector'' models \cite{Arkani-Hamed:1997fq,Luty:1998vr} consider
a class of theories in which DSB can be argued convincingly,
and in which the first two generations of the MSSM arise
as composite states $(P \bar U)$ of a strongly coupled gauge theory.
The fields $\bar U$
acquire large $F$-terms, so that the composites $(P \bar U)$
feel the \susy-breaking directly.  The first and second
generation scalars get large masses, whereas the 
fermion composites remain massless due to chiral 
symmetries.  Since the $\bar U$ fields also carry Standard Model charges,
they communicate \susy-breaking to the rest of the
MSSM through gauge mediation.
The scalar masses for the first and second generation
composite scalars $(P \bar U)$ are parametrically
given in Eq.~(3.7) of \cite{Arkani-Hamed:1997fq} as
$m_\phi^2 \sim \FUb^2/\Ub^2$.
It is convenient in what follows to use the more standard notation
$\FUb \to F$, $\Ub \to M$,
such as appears in~\cite{Giudice:1998bp}.

The messenger scale is the scale of the
strong internal dynamics, corresponding to the
Kaluza-Klein scale \myref{mkk} in the gravitational dual.
Taking into account the parameters chosen in \myref{oset},
the messenger scale is thus
\beq
M = 110 ~ \text{TeV}.
\label{mmes}
\eeq
We will assume $F \approx M$,
as is common in theories where the
messengers couple strongly to the DSB sector.
We also require a large enough $F/M$ in order to
have a viable spectrum, and this too leads
to $F \approx M$.  In particular,
we choose 
\beq
F/M = 90 ~ \text{TeV}.
\eeq
We note that the larger scalar masses in Table \ref{tab:softmUVH}
are somewhat lower than this scale.  This can be explained
by the fact that the localization of the fields is
such that they are in fact a mixture of composite and
elementary modes.  For instance, $\tilde e$ has $b=1.15$,
which as can be seen from Fig.~\ref{m2f} is just
to the IR side of the dashed line that separates
the two localization regimes.

The other ingredient
that is needed to compute the effects of
gauge mediation is the number of messengers,
$N_m$.  In practice we set $N_m=2$, since this gives
rise to an attractive LHC phenomenology and
satisfies the experimental constraints that will be discussed below.
We note that the $B\mu$ term and $A$ terms are generated
radiatively, with the boundary condition that
they vanish at the messenger scale \myref{mmes}.
This is fairly constraining and significantly influences the model parameters.  
In particular, we have adjusted the model to obtain viable electroweak
symmetry breaking and the lightest Higgs mass.

\subsection{Hypercharge FI term constraints}
It is important that a large FI term for hypercharge
is not generated when the heavy first
and second generation scalar fields are integrated 
out~\cite{Dimopoulos:1995mi},\cite{mmssm}.
This amounts to imposing the constraint
\beq
\tr Y \tilde m^2 = 0~,
\eeq
at the messenger scale.
For the leptons, this is not an issue, because
we can have left-right degeneracy for
each generation separately, and $\tr Y = 0$
for $L_i + e_i^c$, $i=1,2,3$.  However for the squarks,
the left-right degeneracy would be broken if all
the hierarchies in the 4D Yukawa couplings were generated
from bulk profiles.  We resolve this potential
difficulty by imposing degenerate $c$'s
for the squarks of the first two 
generations, $Q_i,u_i^c,d_i^c$,
$i=1,2$.  The absence of a one-loop hypercharge
FI term then follows from $\tr Y = 0$ for
each generation of $Q_i,u_i^c,d_i^c$.  The
alternative is to fine-tune the left-right splitting
such that
\beq
\tilde m_{Q_i}^2 - 2 \tilde m_{u^c_i}^2 + \tilde m_{d^c_i}^2 = 0~,
\label{noyc}
\eeq
where $i=1,2$.
However, as we next discuss, FCNC constraints provide another
compelling reason to impose degeneracy amongst the $c$'s,
which is what we will do in all that follows.

\subsection{FCNC's}
\label{fcs}
With all diagonal components of the 5D Yukawa couplings of order one, 
the soft mass matrices $\tilde m_{ij}^2$ that will
arise from the 5D calculation are of a diagonal, nondegenerate
form.  Nondegenerate squarks are very dangerous, and
for this reason we keep the first two generation squarks degenerate
by allowing the hierarchy among 5D Yukawa couplings shown
in Table~\ref{SMfields}.  It is interesting that
the small hierarchy, $m_c/m_u$, in 5D Yukawa couplings
mimics that which occurs in the ``meson'' single-sector models,
where additional dynamics at a high scale was assumed
to generate the necessary ratios.  In addition to
1-2 mixing constraints, the generation 1-3 and 2-3 mixing 
cannot be too large.  Since the corresponding
splittings among the scalars is determined by
the ratio of the Kaluza-Klein scale $\pi z_1^{-1}$ to the gauge
mediation scale $\alpha/(4\pi) F/M$, the
IR scale $z_1^{-1}$ cannot be too large.  Note however
that the wavefunction overlap in our model still 
solves the big fermion mass hierarchies, such as $m_t/m_e$.
In Table~\ref{fcet} we compare a representative example of our 
model to experimental constraints, where the latter are
extracted from the recent results of Ciuchini 
et al.~\cite{Ciuchini:2007ha}.  To obtain the
mixing parameters in the super-CKM basis, it
is necessary to make some assumptions regarding
the quark Yukawa couplings.  The assumption that
reproduces the CKM matrix and leads to tree-level mixing
in the down sector is:\footnote{An analogous
assumption of mixing in the up sector can be
made.  However it leads to weaker constraints.}
\beq
Y^d = V D^d V^T,
\label{por}
\eeq
where $V$ is the CKM matrix and $D^d$ is the
diagonal matrix of down-type quark masses.
This assumption is implemented in the
``off the shelf'' version Softsusy, 
and has been used to generate the ``model''
mixing parameters $|\delta^d|$ that are
shown in Table \ref{fcet}.
Our example model
has been adjusted to sit just at the edge of
the exclusion bound in the most constraining
channel, $b \to s X$.

\begin{table}
\begin{center}
{\footnotesize
\begin{tabular}{|l|l|l|} \hline
$|\delta^d|$ & model & 95\% CL \\ \hline
12/LL & \enot{2.1}{-4} & \enot{1.4}{-2}  \\
12/RR & \enot{2.1}{-4} & \enot{9.0}{-3}  \\
12/LR & \enot{8.5}{-12} & \enot{9.0}{-5}  \\
12/RL & \enot{4.9}{-13} & \enot{9.0}{-5}   \\
13/LL & \enot{2.2}{-2} & \enot{9.0}{-2}  \\
13/RR & \enot{2.1}{-2} & \enot{7.0}{-2} \\
13/LR & \enot{3.6}{-8} & \enot{1.7}{-2}   \\
13/RL & \enot{5.1}{-11} & \enot{1.7}{-2}   \\
23/LL & \enot{1.6}{-1} & \enot{1.6}{-1}  \\
23/RR & \enot{1.6}{-1} & \enot{2.2}{-1}  \\
23/LR & \enot{2.6}{-7} & \enot{4.5}{-3}   \\
23/RL & \enot{6.4}{-9} & \enot{6.0}{-3}   \\ \hline
\end{tabular}}
\caption{Comparison of our model to experimental bounds
on down-type squark mixing parameters, in a standard
notation.  The 1-3 mixings are
constrained by $\Delta m_B$ and $\beta$ measurements,
whereas the 2-3 mixings are constrained by $b \to s X$ and $\Delta m_{B_s}$
measurements.
It can be seen that the latter are the most constraining.  
\label{fcet}}
\end{center}
\end{table}

Leptonic FCNC's, such as those yielding
$\mu \to e \gamma$, can be avoided by assuming
a diagonal lepton Yukawa matrix,
since we do not embed the current theory
in a GUT that would relate quark and lepton mixing.\footnote{In the case
of a non-diagonal lepton Yukawa matrix, the
dominant lepton flavor violating effects would 
come from diagrams involving sleptons and gauginos.  A much
smaller effect would arise from Kaluza-Klein Z-bosons, because they
are quite heavy in this class of models.}
The addition of a right-handed neutrino $\nu^c$
localized near the IR brane in our model would
allow for Dirac neutrino masses from small
effective 4D Yukawa couplings ($Y_\nu \lappeq 10^{-11}$),
thereby avoiding problems of lepton
flavor violation.  This scenario is easily embedded 
into an $SU(5)$ GUT extension of our model.

\subsection{Tachyonic stop/sbottom constraint}
It is well-known that heavy first two generation
squarks and light third generation squarks
can lead to a  tachyonic mass-squared for the
latter under renormalization group evolution \cite{AHM97}.
This effect significantly
constrains the class of models considered here.

Consider the $\beta$-function for
the third generation squark doublet
mass-squared, $d \tilde m_{Q_3}^2/d t$,
where $t$ is the logarithmic scale.\footnote{In the
discussion that follows we rely on the results of \cite{MV93}.
For the numerical checks, we use the two-loop
running of Softsusy.}
We denote by $\Lamlt$ the mass scale of the first two
generation squarks, and assume
that this is much larger than the
gluino mass $M_3$ and the third generation
squark masses.  The one-loop contribution
to the $\beta$-function is:
\beq
\beta_{\tilde m_{Q_3}^2}^{(1)} \approx -\frac{\alpha_s}{4\pi} \frac{32}{3} M_3^2,
\label{1lp}
\eeq
leading to an increase in $\tilde m_{Q_3}^2$ as
one flows to the IR.  On the other hand
at two-loop one has:
\beq
\beta_{\tilde m_{Q_3}^2}^{(2)} \approx \frac{\alpha_s^2}{(4\pi)^2} 
\frac{4 \cdot 32}{3} \Lamlt^2,
\label{2lp}
\eeq
which tends to decrease $\tilde m_{Q_3}^2$ as
one flows to the IR.  For $\Lamlt \gg |M_3|$,
\myref{2lp} dominates over \myref{1lp}, with the
consequence that
a small ($\lappeq 1$ TeV) value of $\tilde m_{Q_3}^2$ at the
messenger scale $M$ will be driven negative 
before reaching the scale $\Lamlt$
where the first and second generation squarks
decouple, provided $\Lamlt$ is sufficiently far below $M$.
In our model it is a good approximation
that $\Lamlt \approx 0.05M = \e M$.  This
is a sufficient separation for the tachyonic
mass-squared to develop, in contrast to
what occurs in models where $\Lamlt \approx M$.
Thus the necessity to have 
$\tilde m_{Q_3}^2(\text{TeV}) > 0$ leads to the constraint:
\beq
\Lamlt \lappeq 6 M_3.
\label{tsb}
\eeq
One concludes that in addition to the 1-3 and 2-3 FCNC constraints,
the first two generation squarks must not be too heavy
relative to the gluino, so as to avoid developing tachyonic masses
for the third generation squarks.

In practice we have performed our
renormalization group evolution 
analysis using Softsusy~\cite{Allanach:2001kg}, where the one- and two-loop
$\beta$-functions are implemented in full detail---including 
Yukawa couplings and mixing---rather 
than with the approximations \myref{1lp}
and \myref{2lp}.  The discussion above
is only meant to give a leading order explanation
of the effect.  Nevertheless, we find
that the bound \myref{tsb} is a good
approximation to the full-fledged results.

\subsection{The particle mass spectrum}
In Table \ref{jot10} we show the complete soft mass spectrum
using the two-loop RGE code Softsusy \cite{Allanach:2001kg},
for the values of the parameters given in (\ref{oset}),
and $\mu < 0$ for the Higgsino mass parameter.
Boundary conditions are imposed at the messenger
scale \myref{mmes}, and the bulk soft masses of
Table \ref{tab:softmUVH} are added in quadrature
to the gauge mediation masses at that scale.
Softsusy automates a self-consistent determination
of the thresholds for the superpartner spectrum,
taking into account one- and two-loop effects.

Note that these are only the masses for the lightest modes, which are
zero modes in the \adsvt\ limit.  The Kaluza-Klein modes are at the ${\cal O}(100)$ 
TeV scale. The heavy first and second generation
scalar masses arising from the bulk 5D calculation represent {\it bona fide} 
nonperturbative masses in the 4D dual theory that are difficult to calculate 
directly in the strongly coupled gauge theory.

The gravitino mass is obtained from the standard formula 
\beq
m_{3/2} = \frac{F}{ \sqrt{3}\,m_P} = 2.35~ \text{eV}.
\eeq
Furthermore, because $\sqrt{F} = \ord{100}~ \text{TeV}$, this
yields a decay length for $\tilde \chi_1^0$ that is
a fraction of a millimeter.  Thus the NLSP
decays inside of a detector, with well-known
collider signatures, as we will discuss in the next section.

\begin{table}
\begin{center}
\begin{tabular}{|c|c|} \hline
$\tilde e_L$, $\tilde e_R$, $\tilde \nu_{eL}$ & 10160, 10150, 10160 GeV \\
$\tilde \mu_L$, $\tilde \mu_R$, $\tilde \nu_{\mu L}$ & 5145, 5130, 5145 GeV \\
$\tilde d_L$, $\tilde d_R$, $\tilde u_L$, $\tilde u_R$ & 5905, 5885, 5970, 5890 GeV \\
$\tilde s_L$, $\tilde s_R$, $\tilde c_L$, $\tilde c_R$ & 5905, 5885, 5970, 5890 GeV \\
\hline
$\tilde g$ & 1615 GeV \\
$\tilde b_1$, $\tilde b_2$, $\tilde t_1$, $\tilde t_2$  & 1354, 1369, 1253, 1369  GeV \\
$\tilde \tau_1$, $\tilde \tau_2$, $\tilde \nu_{\tau L}$ & 511, 630, 633 GeV \\
$\tilde \chi_1^\pm$, $\tilde \chi_2^\pm$ & 478, 593 GeV \\
$\tilde \chi_1^0$, $\tilde \chi_2^0$, $\tilde \chi_3^0$, $\tilde \chi_4^0$
& 288, 480, 511, 598 GeV \\ 
$h^0$, $A^0$, $H^0$, $H^\pm$ & 115, 646, 646, 651 GeV \\
$\tilde G$ & 2.35 eV \\
\hline
\end{tabular}
\caption{Particle mass spectrum of the example single-sector model described in the text.
\label{jot10}}
\end{center}
\end{table}

\section{LHC study}
\label{lhc}
Here we present the results of a preliminary LHC
study of the $pp \to 2\gamma + \METa$ signal
in the example single-sector model we are studying,
summarized in Table \ref{jot10}.
The diphoton signal has been studied as a probe for new physics,
for instance, by the experiments at Tevatron; an example is \cite{cdf1}.
The present study was performed using PYTHIA (version 64.08) \cite{Sjostrand:2006za}.
Subsequent studies using detector simulations
would allow for a refinement of the results
summarized here and would complement
closely related LHC studies \cite{n97079,cr99019}.  Nevertheless, it can be
seen from the results given below
that for the spectrum studied here,
it is easy to remove virtually all backgrounds
and have discovery with 1-10 fb${}^{-1}$ of data.

Due to R-parity, two SUSY particles are produced (except
in the case of Higgs pair production) and
at the end of the decay chain one has:
\beq
2 \tilde \chi_1^0 \to 2(\gamma + \tilde G),
\label{dpevt}
\eeq
where $\tilde G$ is the gravitino.
As a consequence, two very hard photons and abundant missing
transverse energy ($\METa$) characterize the SUSY events.
The decay length for $\tilde \chi_1^0$ is a fraction
of a mm, so the decay occurs inside the tracking system
and will be unobservable.

\subsection{Comparison to conventional gauge mediation}
\begin{figure}
\begin{center}
\includegraphics[width=2.7in,height=4in,angle=90]{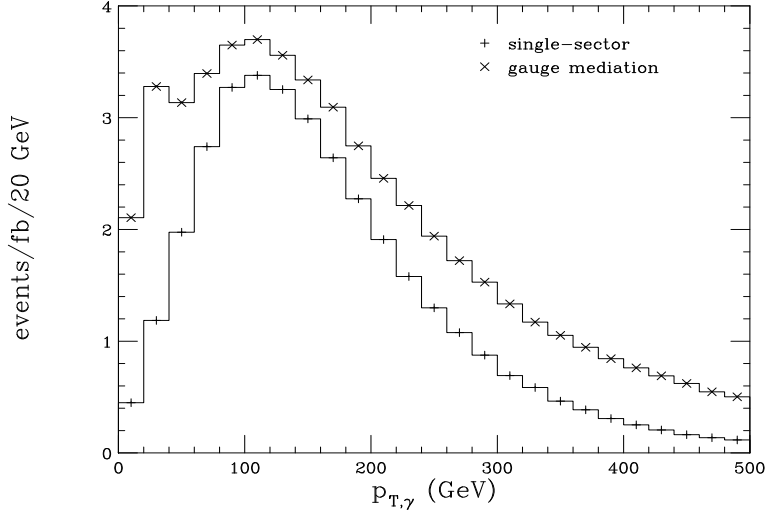}
\caption{The photon differential $p_T$ distribution in our example model, 
compared to a comparable GMSB model, with only nominal cuts (cf.~Sec.~\ref{nrc}).
High $p_T$ bins are especially useful as discriminators,
and would be statistically significant with 10 fb$^{-1}$ of data.
\label{ncr}}
\end{center}
\end{figure}

As can be seen from Fig.~\ref{ncr},
rates for the diphoton events \myref{dpevt} are reduced by a factor of $\sim$50\%
relative to conventional gauge mediation
with the same values of $M, F/M, \tan \beta$,
number of messengers $N_m$ and $\mu < 0$.  
This is just because many sparticles in our model
are beyond LHC reach, due to the large nonpertubative
contribution to their masses (cf.~Table~\ref{tab:softmUVH}).  From the
scaling of the significance $S/\sqrt{B}$, we conclude
that in our model $\sim$4 times more data will be required
for discovery in the diphoton channel compared to GMSB.
Nevertheless, we will show below that it
is easily detectable above backgrounds.
Thus, we can distinguish our model from
GMSB and discover it with 1 to 10 fb$^{-1}$ of
LHC data; i.e., with less than a year of
``well-understood'' data.

\subsection{Signal versus background analysis}
We now turn to the question of whether
the diphoton signal can be seen above
the Standard Model (SM) background at the LHC and show that it will 
indeed be possible. 

\subsubsection{Nominal cuts}
\label{nrc}
We are interested in diphoton events that are reasonably
clean.  That is, the photon should be fairly energetic,
isolated and identifiable.
Isolation cuts are imposed as follows.  An inner
cone of $R_{\text{in}}=0.02$ is defined, centered on the photon.
The size of the cone is based on the CMS ECAL
segmentation of $0.017 \times 0.017$ in $\eta \times \phi$ space.
We compute the total energy $E_{\text{in}}$ of all visible particles in
this cone.  Following \cite{n97079}, we define a wider
cone with $R_{\text{wide}} = 0.3$, again centered on the photon.
Similarly we compute the total energy $E_{\text{wide}}$ of all visible particles in
this cone.  For isolation, we require that at least 90\% of the energy
is contained in the inner cone.  To summarize:
\beq
R_{\text{in}} = 0.02, \quad R_{\text{wide}} = 0.3, \quad
E_{\text{in}}/E_{\text{wide}} \geq 0.9.
\eeq
We also impose a kinematic cut on the photons
such that they are modestly hard:
\beq
p_{T,\gamma} \geq 10 ~ \text{GeV}.
\eeq
Finally, we require that the photons
fall into the central region, where
resolution and identification is optimal:
\beq
|\eta| \leq 2.5~.
\eeq
The solid lines in Fig.~\ref{pt1ln}
show the $p_{T,\gamma}$ and $\METaa$ distributions of diphoton
events that pass the above cuts, 
obtained from the \susy\ hard processes.  To obtain
these results, $5 \times 10^5$ events were simulated.
Bin counts were subsequently rescaled to yield
distributions for 1 fb${}^{-1}$ integrated luminosity.
It can be seen that the diphoton events
generated through the SUSY processes have a very broad $\METaa$
distribution.  This is largely due to the
gravitinos that escape the detector.
As usual, the broad $\METaa$ distribution will grant us
substantial leverage in removing backgrounds
from Standard Model processes.

\subsubsection{Standard Model backgrounds}
The Standard Model produces backgrounds
that would obscure the signal if only
the nominal resolutions cuts are made.
One background is just diphoton production
through Standard Model processes:
\beq
pp \to \{ gg, q\qbar \} \to \gamma \gamma,
\label{bg1}
\eeq
where in the intermediate step we have denoted the partons
from the $pp$ pair that contribute
to the diphoton hard process.
The other backgrounds involve QCD jets ($j$) that fake
photons (mainly due to $j \sim \pi^0$):
\beq
pp \to \gamma ~ j_{\text{fake}}, \quad
pp \to j_{\text{fake}} ~ j_{\text{fake}}.
\label{bg23}
\eeq

These backgrounds have been, for instance, discussed
in \cite{cdf1,n97079}.  We repeat the discussion from \S4 of
\cite{n97079}.  At the LHC, the cross sections for the
three relevant Standard Model events are shown
in Table \ref{smcs}.
\begin{table}
\begin{center}
\begin{tabular}{|c|c|c|c|} \hline
channel & $\gamma \gamma$ & $\gamma j$ & $j j$ \\ \hline
cross section & 0.15 $\mu$b & 0.12 mb & 55 mb \\ \hline
\end{tabular}
\end{center}
\caption{Standard Model cross sections, reproduced
from \cite{n97079}. \label{smcs}}
\end{table}
The ratio of cross sections, when
$\gamma$ is replaced by a jet, is roughly 1:1000.
The probability of a jet to fake an isolated
photon is also about
1:1000.  Thus each of the three channels contributes
at about the same order to background.  A crude estimate
of the total background is therefore $3 \times (pp \to \gamma \gamma)$.
We will take this approach in what follows.
However, an interesting follow-on to this
study would be to simulate the $\gamma j$ and $j j$
events, and check to see what the effect of
the background reduction cuts is on the
corresponding distributions.

\subsubsection{Simulation}
In our background study, we simulate $pp$ collisions
at $\sqrt{s} = 14$ TeV, with the hard process
selection in PYTHIA set to:
\beq
gg, q \qbar \to \gamma \gamma,
\eeq
and then show how to reduce this
background relative to signal with kinematic cuts.

The background was accumulated with 4 runs of $5\times 10^5$
events each.  The runs differed by lower and upper
kinematic cuts on the hard process, as implemented
in PYTHIA through the variables CKIN(3) and CKIN(4),
shown in Table~\ref{smes}.
\begin{table}
\begin{center}
\begin{tabular}{|c|c|c|} \hline
Run & CKIN(3) & CKIN(4) \\ \hline
1 & 0 & 50 \\
2 & 50 & 100 \\
3 & 100 & 250 \\
4 & 250 & $\infty$ \\ \hline
\end{tabular}
\end{center}
\caption{Arrangement of the kinematic
cuts that were made for the estimation of
Standard Model backgrounds. \label{smes}}
\end{table}
Then these were summed, weighted by
the corresponding cross sections measured
in each run.  The reason that this was done is that the low $p_T$ and
$\METaa$ Standard Model events would otherwise statistically
overwhelm the higher bins, and one would not get
a representative sample in the latter.  
The hard process kinematic cuts overcome
this, allowing for reliable background estimates
over several decades.

\begin{figure}
\begin{center}
\begin{tabular}{cc}
\includegraphics[width=2.1in,height=3.1in,angle=90]{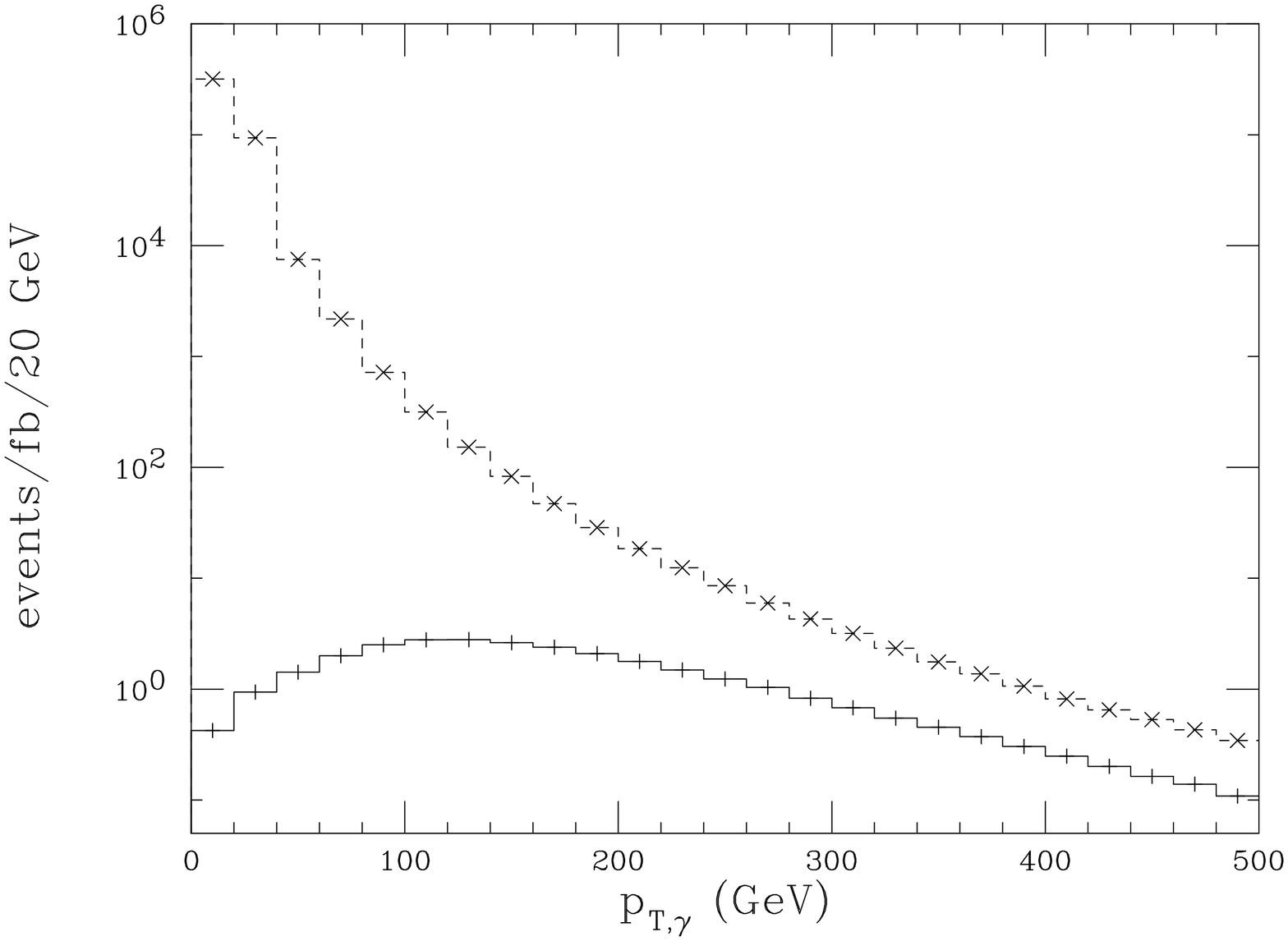} &
\hspace{0.1in}
\includegraphics[width=2.1in,height=3.1in,angle=90]{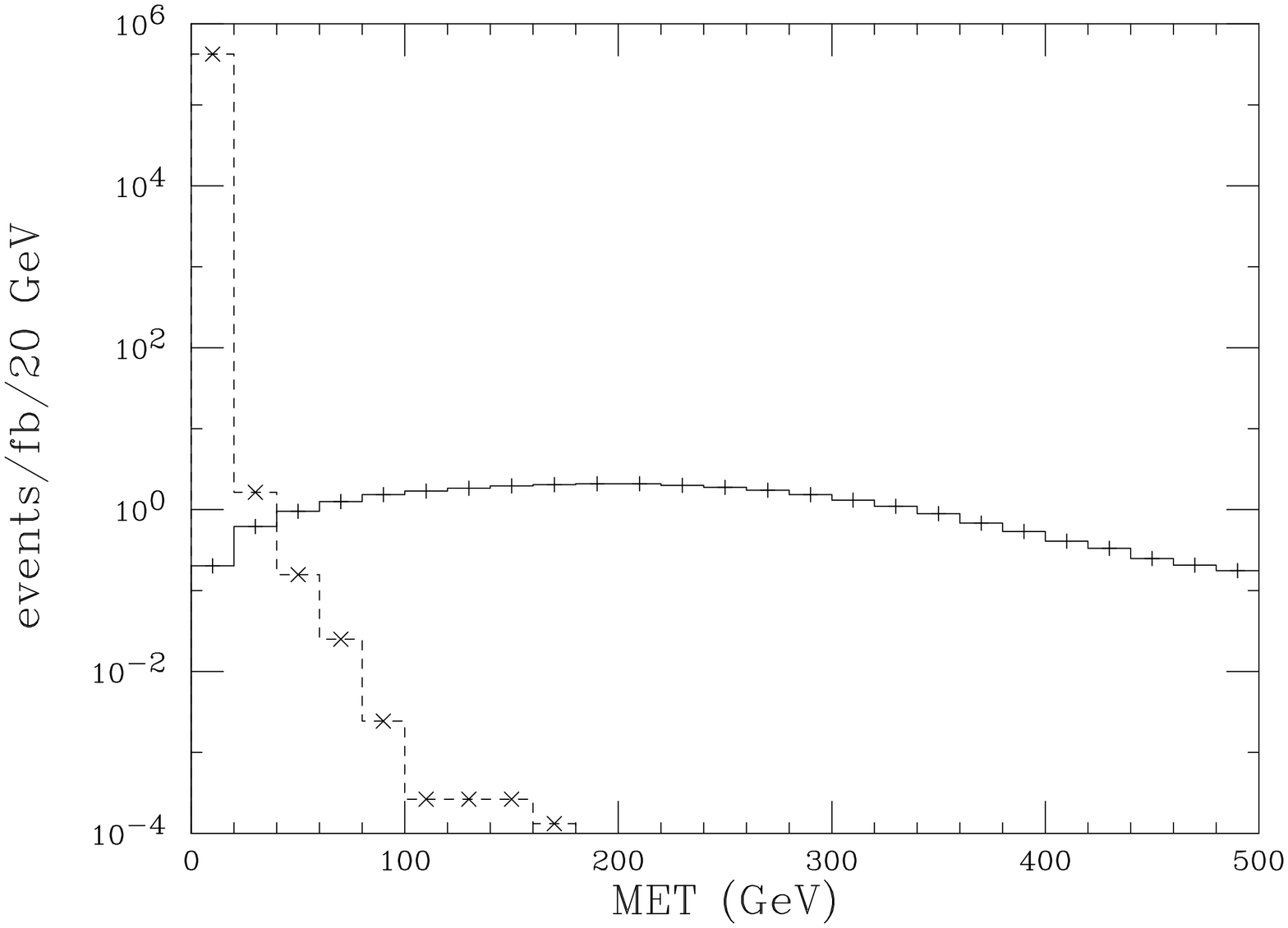}
\end{tabular}
\caption{
Comparison of the single-sector diphoton signal (solid) to 
background (dashed) of Eq.~\myref{bg1}.
Here both the photon $p_T$ and event $\METaa$ distributions are shown.
Only nominal cuts (cf.~Sec.~\ref{nrc}) are made.
\label{pt1ln}}
\end{center}
\end{figure}

The background (dashed) is compared to signal (solid) in Fig.~\ref{pt1ln}.
It can be seen from Fig.~\ref{pt1ln}
that it would be challenging to detect the diphoton
SUSY signal in the $p_T$ distribution,
without further cuts.  (Note that the vertical
axis is a logarithmic scale.) However, Fig.~\ref{pt1ln}
demonstrates that the $\METaa$ distribution would yield
rapid discovery of ``new physics'' 
in the diphoton channel, since in
all but the lowest bin the counts are far in excess of
background.  Furthermore, it is clear that a $\MET$
cut will remove most of the background.

\begin{table}
\begin{center}
\begin{tabular}{|c|c|c|c|}\hline
Integrated & & & \\
Luminosity & SUSY & SM $2\gamma$ & SM $2\gamma$ + fakes \\ \hline
1 fb${}^{-1}$ & 27.6 & 0.0285 & $\lappeq 0.1$ \\
10 fb${}^{-1}$ & 276 & 0.285 & $\lappeq 1$ \\ \hline
\end{tabular}
\caption{Comparison of event rates for
the \dpmet\ channel, after cuts.
Note that the Standard Model (SM) backgrounds
are negligible, and the SUSY signal is spectacular.
\label{ert}}
\end{center}
\end{table}

\begin{figure}
\begin{center}
\begin{tabular}{cc}
\includegraphics[width=2.1in,height=3.1in,angle=90]{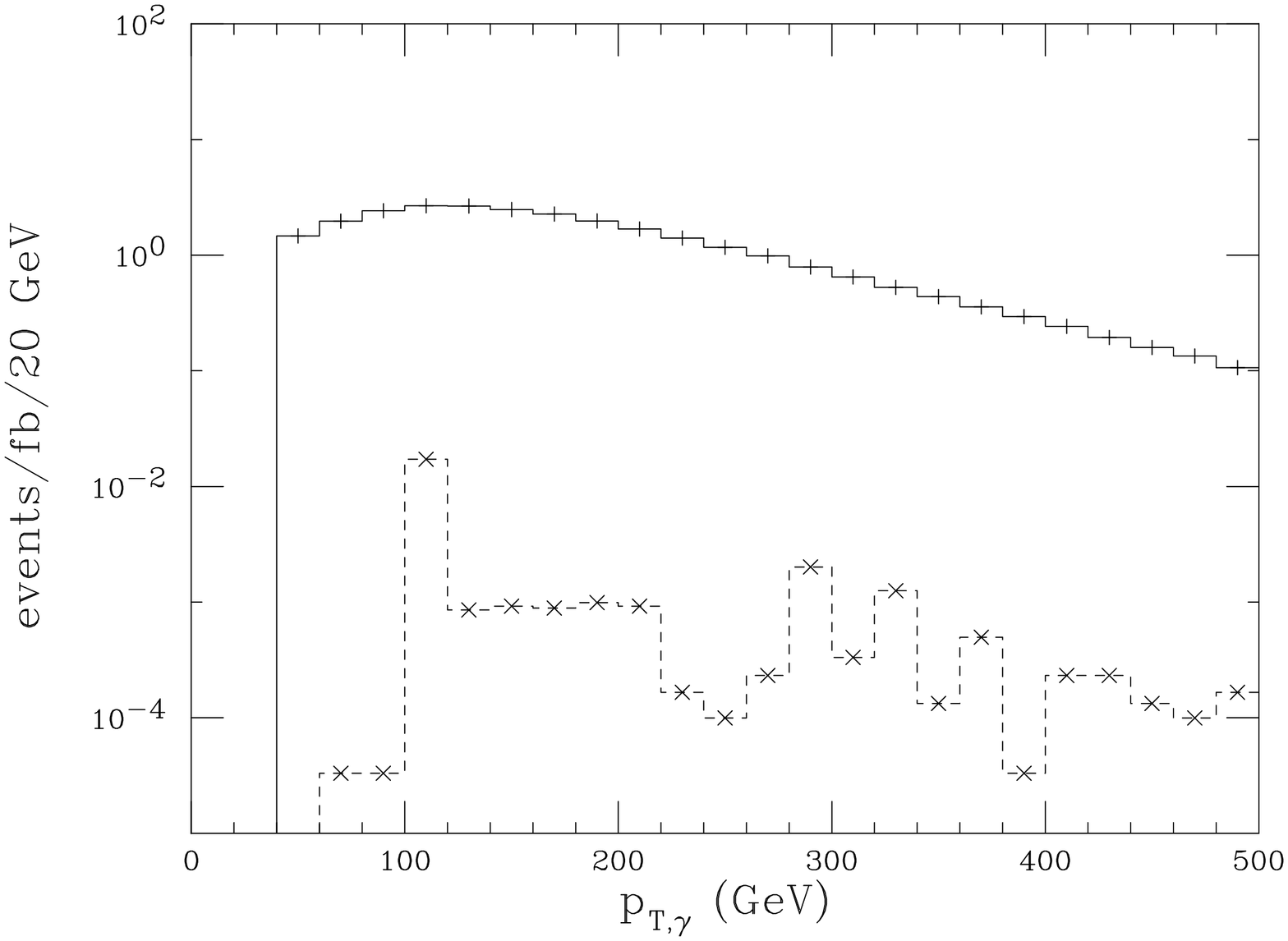} &
\hspace{0.1in}
\includegraphics[width=2.1in,height=3.1in,angle=90]{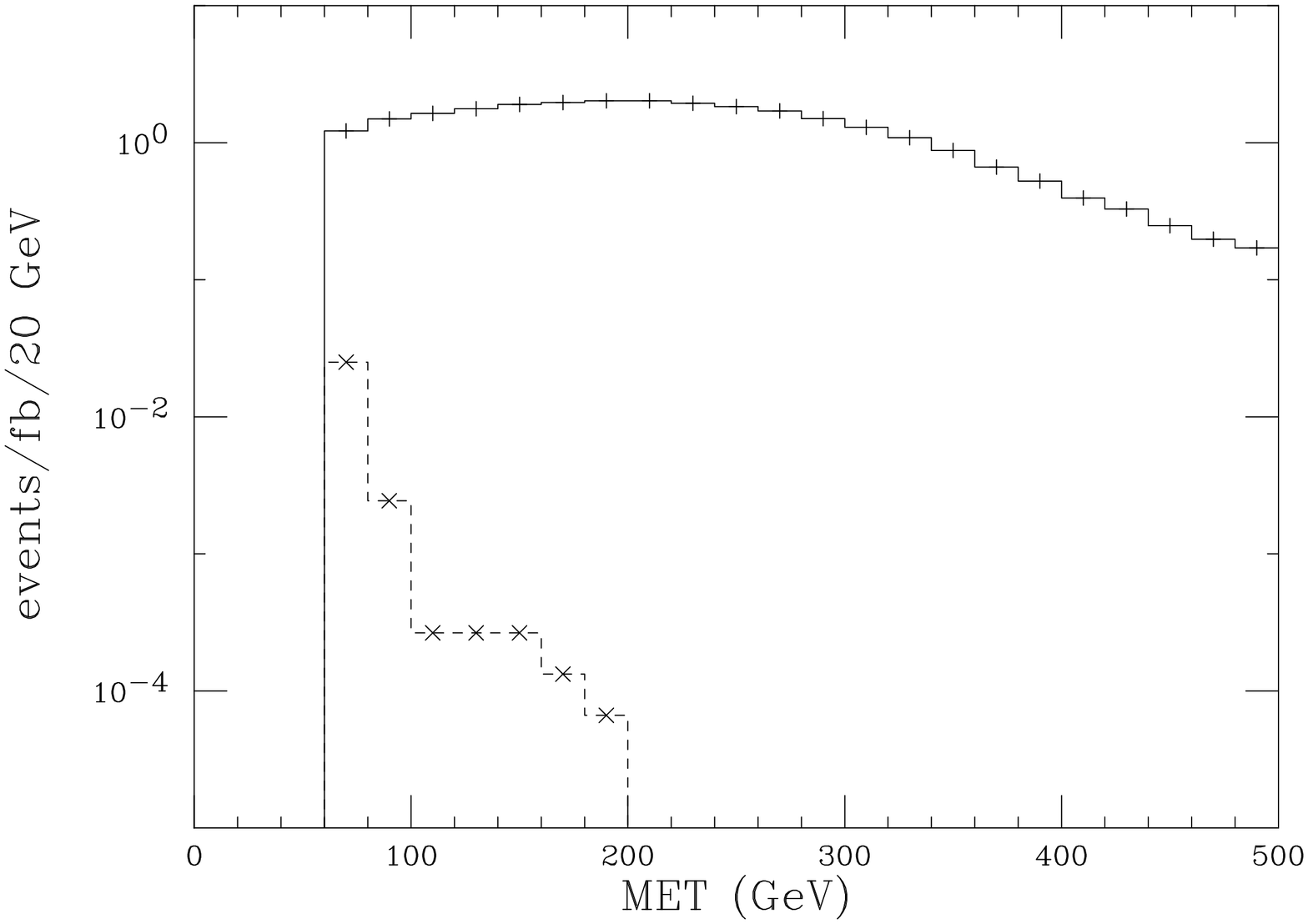}
\end{tabular}
\caption{
Comparison of the single-sector diphoton signal (solid) to background (dashed)
Eq.~\myref{bg1}.
Here both photon $p_T$ and event $\METaa$ distributions are shown.
Cuts to remove background, Eq.~\myref{brc}, have been made,
removing virtually all the background.
It can be seen that the signal will be spectacularly visible.
\label{pt2ln}}
\end{center}
\end{figure}

\begin{figure}
\begin{center}
\begin{tabular}{cc}
\includegraphics[width=2.1in,height=3.1in,angle=90]{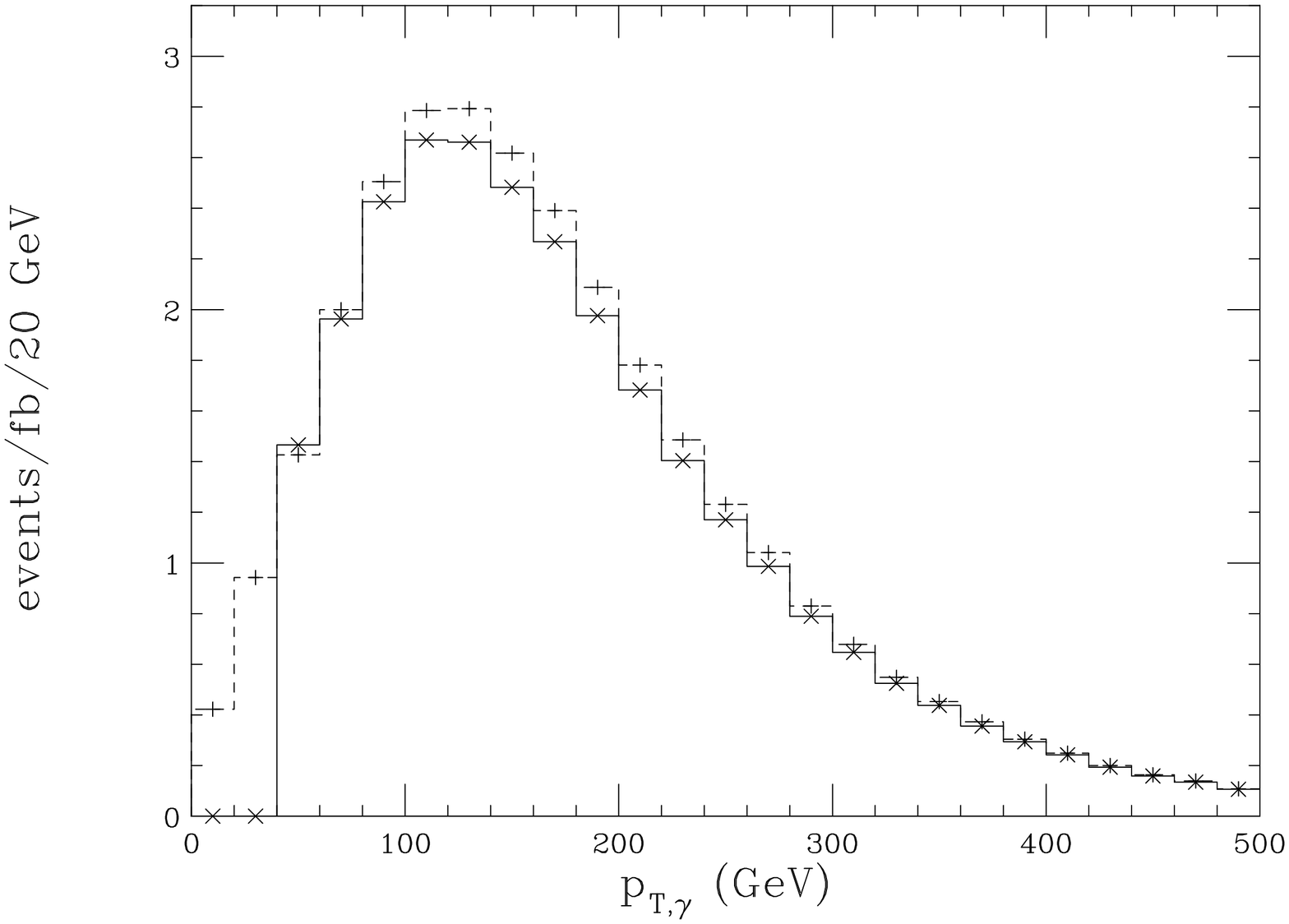} &
\hspace{0.1in}
\includegraphics[width=2.1in,height=3.1in,angle=90]{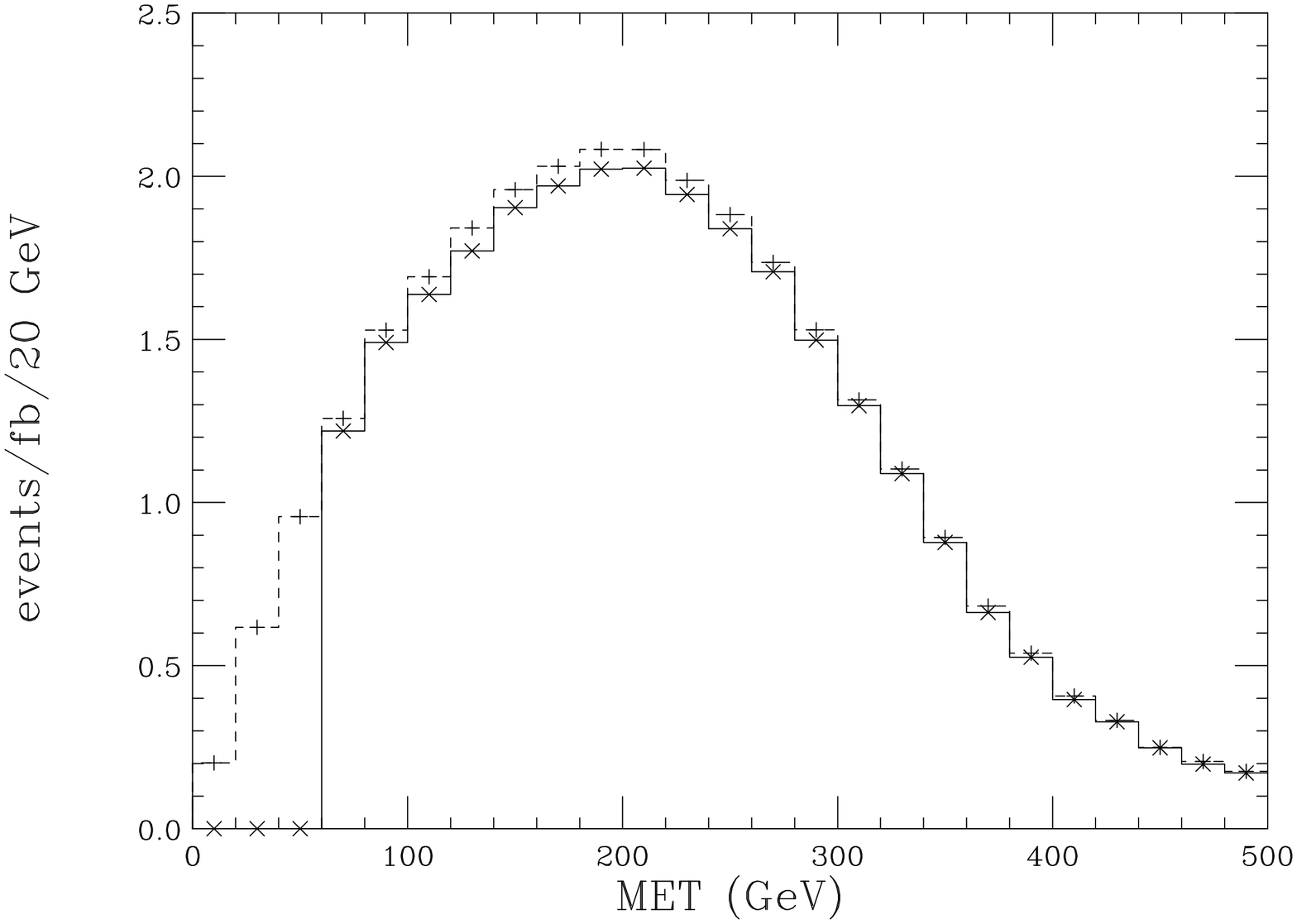}
\end{tabular}
\caption{
Comparison of the single-sector \dpmet\ signal
before (dashed) and after (solid) background reduction cuts.
Distributions of photon $p_T$ and event $\MET$ are shown.
It is clear that not much signal is lost from the cuts \myref{brc}.
\label{pt1_2}}
\end{center}
\end{figure}

\subsubsection{Background reduction cuts}
We impose isolation and central region cuts as in \S\ref{nrc}.
As just mentioned, a $\MET$ cut will remove most of
the background in the diphoton channel.  Furthermore,
Fig.~\ref{pt1ln} shows that the Standard Model
diphoton events are predominantly of low $p_T$.
Based on these results,
we impose the following kinematic cuts to reduce background:
\beq
p_{T,\gamma} \geq 40 ~ \text{GeV}, \quad
\MET \geq 60 ~ \text{GeV} .
\label{brc}
\eeq
The results are shown in Fig.~\ref{pt2ln}.
It can be seen that backgrounds (dashed)
are orders of magnitude smaller than the signal (solid).
With 1-10 fb${}^{-1}$ of data, virtually no
background events occur.  We also note that
the signal is hardly impacted by the cuts \myref{brc}.
This is shown in Fig.~\ref{pt1_2}.

The total event rates after the cuts are shown in
Table \ref{ert}.  We estimate the Standard Model (SM)
``$2\gamma$ + fakes'' rate by approximately three
times the rate obtain from the SM diphoton events,
as explained above.

\subsection{Summary}
The simple $p_{T,\gamma}$ and $\METaa$ cuts \myref{brc} suffice
to remove virtually all SM backgrounds for the \dpmet\ signal.
Discovery of the example model within the first 10 fb$^{-1}$ 
of well-understood data is a certainty, and would occur
during the first few years of the LHC experiment.
Further studies of the jet-fake backgrounds are nevertheless
warranted, because we would like to understand the
exclusion bounds for single-sector models with 
the \dpmet\ signal more generally.  Also, it is important
to determine how $\METaa$ resolution will affect our
signal-to-background results, since poorly measured
background events would end up in the sample after
cuts.  However, because the background is several
orders of magnitude below the signal after our cuts
are imposed, the $\METaa$ resolution should not
pose a difficulty for discovery of the model.

\section{Discussion and Conclusion}
We have presented a 5D dual gravity model of 4D single-sector 
supersymmetry breaking models. These models naturally
explain the scale of supersymmetry breaking and the fermion 
mass hierarchy without invoking a messenger sector.  They lead to 
a distinctive particle spectrum consisting of heavy $({\cal O}(10\,\text{TeV}))$
first and second generation squark and slepton masses. The 
remaining sparticles are lighter $({\cal O}(\text{TeV}))$
so that at low energies only the gluinos, charginos, neutralinos and 
third generation squarks and sleptons will be accessible at the LHC. 
The LSP is the gravitino. This spectrum has been previously studied~\cite{Dimopoulos:1995mi, Pomarol:1995xc}  and is reminiscent of the ``more minimal" 
supersymmetric standard model~\cite{mmssm} .  
The most striking signal at the LHC is from \dpmet, which 
will be easily detectable after 1-10 fb$^{-1}$ of ``well-understood''
data is accumulated.

The dual 4D interpretation of our model
is that the first two generations of fermions and bosons would be composite
states of some strongly coupled gauge theory (``superglue") that is responsible 
for both the scale of supersymmetry breaking via dimensional transmutation and
the fermion mass hierarchy via large anomalous dimensions for fermionic operators
in the gauge theory. The remaining particles are elementary fields that couple weakly
to the composite supersymmetry breaking sector.  This holographic interpretation
is qualitatively identical (i.e., the ``big picture'' is just Fig.~\ref{genmass})
to single-sector models that were explicitly constructed in 
four dimensions~\cite{Arkani-Hamed:1997fq,Luty:1998vr}. Our 
5D model not only has a calculational advantage over 4D strongly coupled gauge theories,
where at best only naive dimensional
analysis estimates are possible, but also uses the AdS/CFT correspondence 
to identify the ratio of the Planck scale to the scale of supersymmetry breaking with the warp factor 
and the fermion mass hierarchy as arising from wavefunction overlap in the bulk.

While we have presented an alternative 5D  interpretation of 4D models of
dynamical supersymmetry breaking, further questions remain that would be interesting
to explore. In particular our 5D model was motivated by type IIB supergravity 
solutions in ten dimensions which are described by nonsupersymmetric backgrounds 
that admit a dual 4D description.  However these explicit supergravity solutions
do not contain the MSSM particle content and therefore the MSSM fields were
introduced by hand in our effective 5D gravity description.
An interesting question to address is how in detail the MSSM 
content could in fact be obtained in the bulk from probe D7 branes~\cite{Gherghetta:2006yq} with
the requisite values of the $c$ parameters~\cite{ABV06}.
Once the mathematical techniques for this part of the theory are fully worked out, 
the field content of the dual 4D description would then be
known, leading to completely defined and fully calculable models.

Our phenomenological model does not solve the little hierarchy problem or  mu problem
associated with the supersymmetric Higgs sector. This is not surprising
since the fifth dimension does not affect this sector at tree
level and we inherit the problem
from the MSSM. However proposed solutions such as including a gauge singlet 
field in non-minimal versions of the MSSM could be straightforwardly added
to our 5D scenario. In addition the fermion sector could easily be extended 
to include neutrino masses via the addition of a right-handed neutrino $\nu^c$. A seesaw 
mechanism is naturally implemented by localizing $\nu^c$ on 
the UV brane or alternatively, Dirac neutrinos could be
obtained from the localization of $\nu^c$ quite near the IR brane,
as discussed in Sec.~\ref{fcs}.
Furthermore it would be interesting to embed our scenario into a grand
unified theory. Extra dimensions have been extremely useful
in providing novel ways to break gauge symmetries and such
mechanisms could be implemented in our scenario.  This 
question will also be relevant for gauge coupling unification
which is essentially the same as in the MSSM provided the ``preons"
of the strongly coupled gauge theory arise in complete SU(5) multiplets.
Nevertheless all these issues deserve further study.

Our scenario also has interesting consequences for cosmology and in
particular dark matter. The $\tilde \chi_1^0$ NLSP will remain in thermal 
equilibrium until its mass scale is reached, with an approximate decoupling 
temperature $T_d \sim m_{\tilde \chi_1^0}/20 \approx 15$ GeV.  Due to the
submillimeter decay length, the neutralino density converts
to gravitino LSP's immediately.  In the class of models
considered here the gravitino mass is $\ord{1}$ eV,
so they will be relativistic down to the temperature
of galaxy formation $T_g \sim 1$ eV.  Their relic
abundance is therefore reduced by a factor $T_g/T_d \sim 10^{-10}$,
so small as to have no effect on large-scale structure.  This is
to say, they are entirely harmless from an astrophysical standpoint.
On the other hand, neither the NLSP nor the
LSP can provide the needed dark matter density. It is an interesting question 
whether or not strongly-interacting
dark matter from the 100 TeV scale gauge theory
could produce dark matter candidates with
the necessary properties. Otherwise either axions or neutrinos could provide 
alternative possibilities.

In summary, we have provided a simple 5D framework in which to study 
single-sector 4D models of dynamical supersymmetry breaking. 
This string-inspired framework has the advantage of providing a mathematical
tool in which to calculate the mass spectrum of the strongly coupled 
4D theory and offers the hope of eventually building a complete model
from the top down in string theory.

\begin{acknowledgements}
JG thanks Keith Olive for helpful discussions on the cosmology of our model and 
Roger Rusack for input on some of the cuts that were used, relative to the CMS 
detector. TG thanks Kaustubh Agashe, Antonio Masiero, and James Wells for useful 
discussions on FCNC constraints. This work was supported in part by a Department of 
Energy grant, No. DE-FG02-94ER40823 at the University of Minnesota.  TG is also 
supported by an award from the Research Corporation.
\end{acknowledgements}

\appendix

\section{Scalar Fields}
\label{sec:ScalarField}

To represent the squarks and sleptons in the bulk 
we first consider a complex scalar field $\Phi(x^{\mu},z)$ in a slice of 
AdS$_5$~\cite{gp}.  The 5D action in the background metric (\ref{defmetric}) reads
\beq
S_{\Phi} = \int d^4 x dz \sqrt{-g} \left( \partial_M \Phi^* 
\partial^M \Phi + M_{\Phi}^2 \Phi^*\Phi \right)~,
\eeq
where $g={\rm det}\,g_{\mu\nu}$. The bulk mass parameter $M_{\Phi}^2$ is given by
\beq
M_{\Phi}^2 \equiv a k^2 + 2bk^2  z \left[\delta(z-z_0)-\delta(z-z_1) \right],
\eeq
where $a$ and $b$ are dimensionless parameters. The equation of motion
\beq
       \partial_{\mu} \partial^{\mu} \Phi + A^{-3} \partial_{5}(A^3 \partial^5 \Phi) -ak^2 A^2 \Phi=0~,
\eeq
is solved by assuming the usual separation of variables $\Phi(x^{\mu},z) = \sum_{n=0}^{\infty} 
\phi_n (x^{\mu}) \tilde{f}_n(z)$. The fields $\phi_n(x^{\mu})$ are the 4D Kaluza-Klein modes 
with profiles $ \tilde{f}_n(z)$ along the extra dimension. A massless zero mode is obtained in the supersymmetric limit $\epsilon\rightarrow 0$ by imposing a modified Neumann boundary condition
$\left(\tilde{f}_n'-bk^2 z A^2\tilde{f}_n\right) \Big\vert_{z_0,z_1} =0$, and a tuning,
 $b=2\pm \sqrt{4+a}$, between the bulk and boundary mass parameters. This leads to the
 massless mode
\beq 
   f_0(z)= \frac{1}{N} (kz)^{b-1}~,
   \label{eq:scalrZM}
\eeq
where $N=\sqrt{(e^{2(b-1)\pi kR} -1)/(k(b-1))}$ is a normalization constant. Note that in (\ref{eq:scalrZM})
we have written the profile with respect to a flat metric $(f_0=(kz)^{-1} \tilde f_0)$ to make manifest the localization properties.
Therefore for bulk masses satisfying the Breitenlohner-Freedman bound $a\geq -4$~\cite{bf},
we have $-\infty< b<\infty$, and the scalar zero mode (\ref{eq:scalrZM})  can be localized
anywhere in the bulk.

In the case of the deformed AdS$_5$ background (\ref{defmetric}) with $\epsilon\neq 0$ the zero 
mode properties will change. Restricting to $\epsilon\ll 1$, the solution for the zero mode profile 
can be written as a perturbation series up to first order:\footnote{The
underlying type IIB supergravity background is only
determined to $\ord{\e}$ in any case; cf.~Sec.~\ref{defbk}.}
\beq
\tilde{f}_0(z) = \tilde{f}_0^{(0)} + \epsilon \tilde{f}_0^{(1)} + \mathcal{O}(\epsilon^2),
\label{zmpf}
\eeq
and accordingly for the zero mode mass squared
\beq
(\tilde{m}_0)^2= (\tilde{m}_0^{(0)})^2 + \epsilon (\tilde{m}_0^{(1)})^2 + \mathcal{O}(\epsilon^2).
\eeq
The zeroth order of course corresponds to the previous solution: 
$\tilde{f}_0^{(0)}= (kz)^b/N$ with $\tilde{m}_0^{(0)}=0$.
Dropping all superscript indices, 
the equation of motion for the $\tilde f_0^{(1)}$ 
term of the zero mode profile \myref{zmpf} reads
\beq
\tilde{f_0}'' -\frac{3}{z}\tilde{f_0}' -\frac{a}{z^2}\tilde{f_0} = 
\frac{k^2}{N}\frac{(6b-a)}{(kz_1)^4}(kz)^{b+2} -\frac{\tilde{m}^2}{N} (kz)^b,
\eeq
where we have used the expansion $A'/A=-1/z-2\epsilon z^3/z_1^4 +\mathcal{O}(\epsilon^2)$.
The general solution for the profile is the combination of a homogeneous and an inhomogeneous part:
\beq
\tilde{f_0}= \frac{(kz)^b} {N_1} + \frac{(kz)^{4-b}} {N_2}
+\frac{(kz)^{b+2}}{8N(b-1)}\left[ \frac{(b-1)(b-10)}{(kz_1)^4}(kz)^2+2\frac{\tilde{m}^2}{k^2} \right],
\eeq
where $N_1$ and $N_2$ are constants.
By expanding the boundary conditions to first order in $\epsilon$ the constant ratio $N/N_2$ 
as a function of $z$ and $\tilde{m}$ can be obtained. The requirement that this 
ratio be equal at the two boundaries leads to the desired expression 
for the mass of the scalar zero mode:
\beq
\tilde{m}_0^2= \epsilon k^2\frac{(b-1)(b+10)}{(kz_1)^4}\frac{e^{2 \pi k R b}-1}{e^{2\pi k R(b-1)}-1} + \mathcal{O}(\epsilon^2).
\label{soft mass}
\eeq
Note that the tuning between the bulk and boundary masses in the supersymmetric limit ($a=b^2-4b$) has been used since any deviation leads to higher order corrections in the masses.

\section{Fermion masses}
\label{fms}
A bulk fermion in a slice of AdS$_5$ is described by a four-component Dirac spinor 
$\Psi(x^{\mu},z)$ whose action reads~\cite{gn},\cite{gp} 
\beq
S_{\Psi} =-i  \int d^4 x dz \sqrt{-g} \left( \bar{\Psi} 
e^M_A \gamma^A D_M \Psi + M_{\Psi} \bar{\Psi}\Psi \right),
\label{psa}
\eeq
where $e^A_M$ is the f\"unfbein, and the covariant 
derivative $D_M=\partial_M + \omega_M$, with $\omega_M$ 
the spin connection. The bulk mass must be odd and is
given by $M_{\Psi}\equiv c k\,\textrm{sgn}(z)$ where $c$ 
is a dimensionless parameter.

Massless zero modes persist in the deformed AdS$_5$ background.  In terms of the
rescaled field $\hat \Psi \equiv A^2 \Psi$, and tangent
space Dirac matrices $\{ \gamma^\alpha, \gamma^\beta \} =
2 \eta^{\alpha \beta}$, the equations of motion are
\beq
0 &=& \( \delta_\alpha^\mu \gamma^\alpha \p_\mu + \gamma_5 \p_z + c A \) \hat \Psi.
\eeq
As usual, we solve by separation of variables:
$\Psi_{L,R}(x^{\mu},z)= \sum_{n=0}^{\infty} 
\psi_{L,R}^n(x^{\mu}){\tilde f}_{L,R}^n (z)$. 
For the zero modes, we set 
$\delta_\alpha^\mu \gamma^\alpha \p_\mu \psi_{L,R}^0=0$, 
and obtain the solutions
\beq
\tilde f_{L,R}^0 & = & N_{L,R}^{-1} A^{-2}(z) 
\exp \( \mp c \int_{z_0}^z dz' ~ A(z') \)~.
\eeq
Hereafter we drop the superscript and discuss
only zero modes.

The background deformation has not lifted
the fermion zero mode because its mass is
protected by chiral symmetry.  In particular,
this implies that the action \myref{psa} in the
deformed background leads to massless gauginos.
(However, another contribution to the gaugino
action arises in the underlying type IIB supergravity
background; cf.~Appendix \ref{fluxr}.)
To obtain chiral zero modes we impose a $Z_2$ projection
in the usual manner, introducing a 5D Dirac 
fermion $\Psi$ for each 4D Weyl fermion of the MSSM.

To obtain the zero mode profile, we switch to a flat 5D coordinate via
\beq
dy = A(z) dz \quad \Rightarrow
\quad y = \int_{z_0}^z A(z') dz' 
\approx k^{-1} \[ \ln (k z) - \frac{\e z^4}{8 z_1^4} \].
\label{ydf}
\eeq
We also occasionally use the notation $ \pi k R \equiv \ln (k z_1)$.
The zero mode action for the 4D field $\psi_{L,R}(x)$ is then
\beq
\int d^4x dy ~ A^3(y) \tilde f_{L,R}^2(y) \psib_{L,R}(x) 
\delta_\alpha^\mu \gamma^\alpha \p_\mu \psi_{L,R}(x)~,
\eeq
leading to a profile $f_{L,R} \equiv A^{3/2} \tilde f_{L,R}$.
Thus we obtain:
\beq
f_{L,R} &=& N_{L,R}^{-1} A^{-1/2}(z) \exp\( \mp c_{L,R} \int_{z_0}^{z} dz' ~ A(z') \)~.
\label{fermprofile}
\eeq
Expanding in the small parameter $\e$ we find
that the profile is virtually unchanged from
what occurs in the $\e \to 0$ limit:
\beq
f_{L,R} \approx
N_{L,R}^{-1} z^{\half \mp c_{L,R}} \[ 1 + \frac{\e z^4}{4 \zir^4} 
\( 1 \mp \half c_{L,R} \) \].
\label{expp}
\eeq
The normalization constants are\footnote{As usual, one
integrates over $(k^2z_1)^{-1} \leq z \leq z_1$, or
equivalently over $-\pi R \leq y \leq \pi R$,
to take into account the $Z_2$ orbifold.}
\beq
N_{L,R}^2 = \frac{e^{2(1/2\mp c_{L,R})\pi kR}-1}{k(1/2 \mp c_{L,R})}
\[ 1 + \ord{\e} \].
\label{eq:NLR}
\eeq
Again the wavefunction \myref{expp} is written 
with respect to the flat coordinate $y$
to make the localization
properties manifest.  We conclude that a left-handed fermion 
zero mode is UV-localized for $c_L>1/2$ 
and IR-localized for $c_L<1/2$, while a 
right-handed one is UV-localized for $c_R<-1/2$ and 
IR-localized for $c_R>-1/2$.  Note that the corresponding 
chiral partner of the zero mode $f_{R,L}=0$
since the $Z_2$ symmetry requires these fields to vanish.

\section{Yukawa couplings}
\label{ycsect}
The 4D Yukawa couplings result from a wavefunction overlap of the bulk SM fermions with the 
Higgs field~\cite{gp, review}. Each SM fermion is identified with the zero mode of the corresponding
5D Dirac spinor.  A Yukawa term in the 5D 
action with brane-localized up- and down-type Higgs fields
is given by
\beq
\int d^4x dz \sqrt{-g} Y^{5D}_{\psi} 
\frac{H(x)}{N_H} kz\delta(z-z_*) \bar\Psi_{R}(x,z) \Psi_{L}(x,z),
\eeq
for a Higgs fields localized at $z=z_*$.  Here,
$Y^{5D}$ is a 5D dimensionful Yukawa coupling parameter,
$\Psi_{L,R}(x,z)$ is the 5D spinor that contains
an SU(2)$_L$ doublet (singlet) of the MSSM as its zero mode,
and $H(x)$ represents the appropriate Higgs field, $H_u$ or $H_d$.

After Kaluza-Klein decomposition and integration over the extra dimension, the part concerning the zero modes reduces to
\beq
\int d^4x  \frac{Y^{5D}_{\psi} }{N_H N_L N_R} (kz_*)^{-c_L +c_R} H(x) \bar\psi_{R}(x)
\psi_{L}(x),
\eeq
from which we read the effective 4D Yukawa coupling:
\beq
\label{eq:Y(c)}
Y_{\psi} = \frac{Y^{5D}_{\psi} }{N_H N_L N_R}(kz_*)^{-c_L +c_R}.
\eeq
The Higgs normalization constant $N_H$ is found by requiring canonical normalization
of the 4D kinetic term
\beq
S_H^{kin} &=& \int d^4x dz \sqrt{-g} \frac{kz}{N_H^2} \delta(z-z_*) g^{\mu\nu}\partial_{\mu} H^*\partial_{\nu} H,
\eeq
which implies $N_H  = 1/(kz_*)$. The constants $N_L$ and $N_R$ are given in (\ref{eq:NLR}).
For UV-confined Higgs fields, $kz_* = kz_0 = 1$.

\section{Aspects of the underlying supergravity}
\label{sugra}
In subsection \ref{defbk}, we describe how the
deformed background \myref{defmetric} emerges
from the non-\susy\ solution given by Kuperstein 
and Sonnenschein (KuSo) \cite{Kuperstein:2003yt}.
This solution is a perturbation of the 
Klebanov and Strassler (KS) \cite{Klebanov:2000hb}
background, governed by a small parameter $\delta$.
It is based on techniques for solving the type IIB supergravity
equations of motion in the KS context that were
developed in \cite{Borokhov:2002fm}.  In subsection
\ref{fluxr}, we describe how $U(1)_R$ is broken
by the flux background of the KuSo solution, and
how it leads to gaugino masses in the bulk.

\subsection{The deformed background}
\label{defbk}
We will be interested in the KS metric described by
\beq
ds^2 &=& 2^{1/2} 3^{3/4} \bigg[ e^{-5q+2Y} dx_\mu dx^\mu + \frac{1}{9}
e^{3q-8p}(d\tau^2+g_5^2) \nnn && + \frac{1}{6} e^{3q+2p+y} (g_1^2 + g_2^2)
+ \frac{1}{6} e^{3q+2p-y} (g_3^2 + g_4^2) \bigg].
\label{ksgf}
\eeq
It is parameterized by
\beq
q,Y,p,y,
\label{tdfs}
\eeq
which are all functions of the radial coordinate $\tau$.
This is related to the $z$ coordinate
through $z \sim e^{-\tau/3}$.  It is important
in what follows that the boundaries of our space, $z_0$ and $z_1$,
are both at $\tau \gg 1$.  Smaller values of $\tau$ have
been integrated out and replaced by an effective IR
brane, as described in \cite{Gherghetta:2006yq}.
For this reason, $e^{-\tau/3}$ can be treated
as a small parameter in the manipulations
that we now summarize.

In the KuSo background one has KS plus a small deformation.
For instance,
\beq
q = q_0 + \delta \cdot \bar q,
\eeq
where $\delta$ parametrizes the deviation from the KS solution.
A similar notation is introduced for the other three functions
in \myref{tdfs}.
Taking into account the various $\tau \gg 1$ asymptotic
forms of the functions \myref{tdfs} that are given in KuSo, we find that
the metric becomes, in this limit:
\beq
ds^2 &=& h_0^{-1/2} [1 + \delta (-5 \bar q + 2 \bar Y) ] dx_\mu dx^\mu
+ \frac{\e_{KS}^{4/3}}{6 K_0^2} h_0^{1/2} [ 1 + \delta (3 \bar q - 8 \bar p) ]
(d\tau^2+g_5^2)  \nnn &&
+ \frac{\e_{KS}^{4/3}}{2} h_0^{1/2} K_0 [1+\delta(3\bar q+2\bar p+\bar y)]
\sinh^2 \frac{\tau}{2} (g_1^2 + g_2^2)\nnn &&
+ \frac{\e_{KS}^{4/3}}{2} h_0^{1/2} K_0 [1+\delta(3\bar q+2\bar p-\bar y)]
\cosh^2 \frac{\tau}{2} (g_3^2 + g_4^2)~,
\label{ksga}
\eeq
where $h_0$ and $K_0$ are the well-known functions that
appear in the KS solution and $\e_{KS}$ is the parameter
that describes the deformed conifold of the KS solution.

Furthermore, one has from KuSo that the leading
terms in each of the functions are
\beq
\bar q \approx - \frac{2}{5} \mu \tau e^{-4\tau/3}, \quad
\bar Y \approx \half \mu \tau e^{-4\tau/3}, \quad
\bar p \approx \frac{3}{5} \mu \tau e^{-4\tau/3}, \quad
\bar y \approx \mu \tau e^{-\tau/3}.
\eeq
The parameter $\mu$ is given by
\beq
\mu =  2^{10/3} 3g_s M \ell_s^2 X \e_{KS}^{-8/3}~,
\eeq
where $g_s$ is the string coupling, $M$ is the number of
fractional D3 branes, $\ell_s$ is the string
length and $X$ is an integration constant in the KuSo
solution that could be set to unity through a redefinition
$\delta \to \delta/X$.
Continuing with the expansion in powers of the
small parameter $e^{-\tau/3}$, one finds
\beq
ds^2 &\approx& \frac{2^{5/6}}{\sqrt{3 \alpha \tau} } e^{2 \tau /3} 
\[ 1 + 3 \delta \mu \tau e^{-4 \tau /3} \] dx_\mu dx^\mu
+ \sqrt{ \frac{ \alpha \tau }{3} } 2^{-5/2} \e_{KS}^{4/3}
\[ 1 - 6 \delta \mu \tau e^{-4 \tau /3} \]
( d\tau^2+g_5^2 ) \nnn
&& + 2^{-7/2}\sqrt{3\alpha\tau}\, \e_{KS}^{4/3} \[
\( 1 + \delta \mu \tau e^{-\tau /3} \)  (g_1^2 + g_2^2)
+ \( 1 - \delta \mu \tau e^{-\tau /3} \)  (g_3^2 + g_4^2) \] ~,
\eeq
where $\alpha = 4 (g_s M \ell_s^2)^2 \e_{KS}^{-8/3}$.

Finally, we restrict our attention to modes that have a trivial dependence on
the angular coordinates of the compact space ($T^{1,1}$ in the $\tau \gg 1$
limit), represented
here by the forms $g_i$, $i=1,\ldots,5$.  (Modes with
a nontrivial dependence on the angular coordinates
of the compact space will be excitations with
mass of the order the Kaluza-Klein scale, and are therefore beyond
the reach of LHC physics that we study.)  With that
assumption, we arrive at the effective 5D metric
\beq
ds^2 &\approx& \frac{2^{5/6} e^{2 \tau /3} }{\sqrt{3 \alpha \tau} }
\[ 1 + 3 \delta \mu \tau e^{-4 \tau /3} \] dx_\mu dx^\mu
+ \frac{ \sqrt{\alpha \tau} }{4 \sqrt{6}} \e_{KS}^{4/3}
\[ 1 - 6 \delta \mu \tau e^{-4 \tau /3} \]  d\tau^2.
\label{5dre}
\eeq

Because $\tau \sim \ln z$, the powers of $\tau$ that appear in
\myref{5dre} are slowly varying relative to the powers of
$e^{-\tau/3} \sim z$.  On this basis we approximate the
powers $\tau^p$ by a constant $\tau_0^p$, which leads to
a great simplification of the expressions that follow.
The analysis in the main text is made significantly simpler
by this approximation as well.  Nevertheless, we capture
the dominant \susy\ breaking effects of the deformed background,
which is our main intent.

With this in mind, we substitute
\beq
z &=& \frac{3}{2^{5/3}} \sqrt{\alpha \tau_0}\, \e_{KS}^{2/3}\, 
e^{-\tau/3} \( 1 - \frac{9}{10} \delta \mu \tau_0 e^{-4 \tau / 3} \), \nnn
\frac{\e}{z_1^4} &=& - \frac{2^{23/3}}{135 \tau_0 \alpha^2} \delta\, \mu\,
\e_{KS}^{-8/3},\nnn
\frac{1}{k^2} &=&
\(\frac{27 \tau_0}{8}\)^{1/2}g_sM \ell_s^2,
\eeq
into Eq.~\myref{5dre} where $\e_{KS}^{2/3}$ has dimensions of length.  To $\ord{\e}$ 
we obtain from \myref{5dre} the deformed metric given in Eq.~\myref{defmetric},
as is easily verified.

\subsection{Flux breaking of $U(1)_R$}
\label{fluxr}
An important issue is the origin of $U(1)_R$ breaking
in the 10D supergravity description.  Since the gaugino
masses occur at one-loop in the 4D dual gauge theory,
they should be evident at tree level in the supergravity.

Fermionic terms in the supergravity action of D3-branes
have been considered for instance in \cite{Grana:2002tu}.
For the D7-branes that we expect the flavor fields
to come from, the features that we now discuss
should be the same, since they are understood in
terms of a dimensional reduction of 10D fermions
and their couplings to the closed string modes
of type IIB supergravity.  The important result is in Eq.~(9)
of \cite{Grana:2002tu}.  There is a gaugino coupling
to the type IIB supergravity 3-form $G_{(3)}$
\beq
G_{ijk} \lambda \lambda + \hc,
\label{gft}
\eeq
where we use a more standard notation and denote
gauginos as $\lambda$.  Here, $G_{ijk}$ is the (3,0)
holomorphic component of $G_{(3)}$, which only
has ``legs'' in the compact space that is orthogonal
to the 5D space we are reducing to.

In the KuSo background that we study, $G_{(3)}$ is nonvanishing.
Furthermore, there is a $U(1)_R \to Z(2)_R$ breaking corresponding
to\footnote{Further details may be found in \cite{Loewy:2001pq,Kuperstein:2003yt}.}
\beq
\int_{X_3} G_{(3)} \approx c'_1 \delta \mu e^{-\tau/3} 
+ ( c'_2 + c'_3 \delta \mu ) e^{-\tau}
\approx c_1 \e \frac{z}{k^2 z_1^4} + c_2 k^4 z^3 
+ c_3 \e \frac{z^3}{z_1^4},
\label{g3i}
\eeq
where $c_{1,2,3}$ depend at most logarithmically on $z$.
The three dimensional space $X_3$ that is integrated over is
the one that describes the embedding of
the D7-branes into the compact space, which
is $T^{1,1}$ in the UV.  The explicit form of
this embedding, which we leave arbitrary here,
will determine the constants $c_{1,2,3}$.
According to the
the AdS/CFT dictionary, the coefficients of $z^3$ correspond to a
nonvanishing $SU(N)$ gluino condensate
of the strongly coupled 4D gauge theory, and
the coefficient of $z$ corresponds to a
mass for that gluino, due to \susy\ breaking
of the bulk.  These terms provide a source of tree level
$U(1)_R$ symmetry breaking, and hence
MSSM gaugino masses.
Taken together with the fermionic terms of the
form \myref{gft}, we thus arrive at gaugino mass terms
\beq
\Delta S = \int d^4 x \,dz \sqrt{-g}
\( c_1 \e \frac{z}{k^2 z_1^4} + c_2 k^4 z^3 
+ c_3 \e \frac{z^3}{z_1^4} \)
\lambda_i \s^3_{ij} \lambda_j.
\eeq
Taking into account the gaugino profiles,
the deformed metric, and the $\lambda \to g \lambda$
rescaling to obtain a canonical kinetic term (here $g$ is
the gauge coupling corresponding to $\lambda$ and in the
considerations here one begins in the basis where all
components of a vector multiplet have $1/g^2$ as
a prefactor of their kinetic terms), one finds
\beq
m_\lambda \approx \frac{g^2}{N_L^2} \[
\frac{c_2}{2} \( 1 - \frac{5}{24} \e \) (k z_1)^2
+ \frac{c_1 \e}{(k z_1)^4} \ln (k z_1) 
+ \frac{c_3 \e}{2 (k z_1)^2} \]~.
\eeq
It can be seen that all but the first term
are completely negligible.
To obtain a mass that agrees with the one
found at one loop in the 4D dual requires
that $c_2 \approx 1/(kz_1)^3$.  This presumably
has to do with an embedding of the D7 branes
that is supersymmetric in the $\e \to 0$ limit,
and is consistent with the background studied here.
For instance, recall that an integral over $X_3$ is performed
in \myref{g3i}.  It is a property of both the KS and KuSo
backgrounds that the $\int_{S^3} G_{(3)} = 0$ for
the $\tau \sim \ln z$ dependendent part of $G_{(3)}$.
Thus the smallness of $c_2$ could arise from an embedding
that wraps a 3-sphere except in the immediate
vicinity of the IR brane, as in \cite{Gherghetta:2006yq}.
It would be interesting to study this issue
further, though it is beyond the scope of
the present article.  The main point is
that there is a tree level source of $U(1)_R$
symmetry breaking that arises from the $G_{(3)}$
flux background, and that there is a plausible
5D dual for what is found in the 4D gauge theory.

\end{document}